\documentclass[12pt,preprint]{emulateapj}

\def\be{\begin{equation}}
\def\ee{\end{equation}}
\def\bea{\begin{eqnarray}}
\def\eea{\end{eqnarray}}

\usepackage{graphicx}
\usepackage{soul}
\usepackage{color,txfonts}
\usepackage{hyperref}

\begin{document}

\title{GRB/GW association: Long-short GRB candidates, time-lag, measuring gravitational wave velocity and testing Einstein's equivalence principle
}
\author{Xiang Li$^{1,2}$, Yi-Ming Hu$^{3,4}$, Yi-Zhong Fan$^{1}$, and Da-Ming Wei$^{1}$
}
\affil{
$^1$ {Key Laboratory of dark Matter and Space Astronomy, Purple Mountain Observatory, Chinese Academy of Science, Nanjing, 210008, China.}\\
$^2$ {University of Chinese Academy of Sciences, Yuquan Road 19, Beijing, 100049, China.}\\
$^3$ {Max Planck Institute for Gravitational Physics, Albert Einstein Institute, Callinstra\ss e 38, D-30167 Hannover, Germany.}\\
$^4$ {Tsinghua University, 30 Shuangqing Rd, Beijing, 100084, China.}\\
}
\email{Email: yzfan@pmo.ac.cn (YZF)}

\begin{abstract}
Short duration GRBs (SGRBs) are widely believed to be powered by the mergers of compact binaries, like binary neutron stars or possibly neutron star-black hole binaries. Though the prospect of detecting SGRBs with gravitational wave (GW) signals by the advanced LIGO/VIRGO network is promising, no known SGRB has been found within the expected advanced LIGO/VIRGO sensitivity range for binary neutron star system.
We find, however, that the two long-short GRBs (GRB 060505 and GRB 060614) may be within the
horizon of advanced GW detectors.
In the upcoming era of GW astronomy, the merger origin of some long-short GRBs, as favored by the macronova signature displayed in GRB 060614, can
be unambiguously tested. The model-dependent time-lags between the merger
and the onset of the prompt emission of GRB are estimated. The comparison of such time-lags between model prediction to the real data expected in the era of the GW astronomy would be helpful in revealing the physical processes taking place at the central engine (including the launch of the relativistic outflow, the emergence of the outflow from the dense material ejected during the merger and the radiation of gamma-rays).
{We also show that the speed of GW, with or without a simultaneous test of the Einstein's equivalence principle, can be directly measured an accuracy of $\sim 3\times 10^{-8}~{\rm cm~s^{-1}}$ or even better in the advanced LIGO/VIRGO era.}
\end{abstract}
\keywords{gamma-ray burst: general---stars: black holes---stars: neutron---binaries: general}

\section{Introduction}
The  coalescence of a binary compact object
system (either a neutron-star (NS) binary or a stellar-mass
black hole (BH) and NS binary) has been widely suggested to account for short-duration gamma-ray burst (SGRB)
events \citep{Eichler1989,Narayan1992,Nakar2007,Berger2014} that lasted typically shorter than 2 seconds in soft $\gamma-$ray band
\citep{Kouveliotou1993}. Since 2006, it had been suspected that the compact object mergers could also produce the so-called long-short GRBs (also known as the supernova-less long GRBs which are apparently long-lasting but do not show any signal of supernovae down to very stringent limits) which share some properties of both long- and short-duration GRBs \citep{Gehrels2006,DellaValle2006,Gal-Yam2006,Zhang2007}. The compact binary coalescence (CBC) is generally expected to be strong gravitational wave (GW) radiation source and are prime target for some gravitational detectors like advanced LIGO/VIRGO \citep[][see also the latest LSC-Virgo white paper at https://dcc.ligo.org/LIGO-T1400054/public]{Abadie2015,Acernese2015,
Belczynski2010,Belczynski2016}. {On September 14, 2015 the two detectors of the Laser Interferometer Gravitational-Wave
  Observatory (LIGO) simultaneously detected a transient gravitational-wave signal from the merger of two black holes (GW150914, \citet{Abbott2016}). GW150914 is the first direct detection of gravitational waves and the first identification of a binary black hole merger \citep{Abbott2016}. Surprisingly, the Fermi Gamma-ray Burst Monitor (GBM) observations at the time of GW150914 claimed a detection of a weak
gamma-ray transient (i.e., GBM transient 150914) 0.4 s after GW150914 with a false alarm probability of 0.0022 \citep{Connaughton2016}. If true, this is the first GW/SGRB association (see however Savchenko et al. 2016 for some arguments). \citet{Li2016} compared GBM transient 150914 with other SGRBs and found that such an event is remarkably different in the prompt emission properties. The binary black hole merger origin as well as its ``distinguished" prompt emission property suggest that GW150914/GBM transient 150914 is not a typical GW/SGRB association.}

In the absence of successful detection of the gravitational radiation triggered by a ``normal" (S)GRB\citep{Abadie2012,Aasi2014a,Aasi2014b}, a ``smoking-gun" signature for the compact-binary origin would be the detection of the so-called Li-Paczynski macronova (also called a kilonova), which is a near-infrared/optical transient powered by the radioactive decay of $r-$process material synthesized in the ejecta that is launched during the merger event \cite[e.g.,][]{Li1998,Kulkarni2005,Metzger2010,Barnes2013}.
The identification of macronova candidates in the afterglows of the canonical short event GRB 130603B \citep{Tanvir2013,Berger2013}, the long-short burst GRB 060614 \citep{Yang2015,Jin2015} and the short event with extended X-ray emission GRB 050709 \citep{Jin2016} are strongly in support of the CBC origin of some GRBs. A conservative estimate of the macronova rate favors a promising detection prospect of the GW radiation by the (upcoming) advanced LIGO detectors \citep{Jin2015}. We anticipate that in the near future many compact-object-merger driven GW sources would be detected \citep{Abadie2010} and a small fraction of such events would be accompanied by supernova-less GRBs (including both the short and long-short events).

The observation of a ``nearby" supernova-less GRB provides a reliable estimate of
the time, sky location and distance of a potential binary merger signal. This significantly reduces the
parameter space of a follow-up GW search and consequently
could be used to reduce the effective detection threshold and effectively increase the detectors' sensitivity and their detection rate \citep[e.g.,][]{Kochanek1993,Finn1999,Harry2011,Kelley2013,Nissanke2013,Dietz2013,Williamson2014,Clark2015,Bartos2015}.
In this work we examine whether
some SGRBs and/or long-short GRBs are within the horizon of the advanced LIGO/VIRGO network
and discuss the model-dependent time lag between the coalescence and GRBs.

This work is structured as follows: In Sec. 2 we discuss/summarize the prospect of detecting GW-associated GRBs in the era of advanced LIGO/VIRGO and examine whether any recent GRBs (either SGRBs or long-short GRBs) are within the horizon
of the advanced LIGO/VIRGO network. In Sec. 3 the  model-dependent time lags between the merger and GRBs are presented and the possibility of revealing the nature of merger remnants with such time delays is discussed. The expected progress in measuring the speed of GW with the future data are investigated in Sec. 4. Our results and discussion are presented in Sec. 5.

\section{The prospects of detecting GW signal associated SGRBs and long-short GRBs}

\subsection{The prospect of detecting SGRBs associated with GW signals}\label{sec:SGRB-prospect}
The strategy of the targeted search for GWs associated with short GRBs has been extensively discussed in \citet{Harry2011} and
\citet{Williamson2014}. The prospect of detecting SGRBs with GW signals has also been widely estimated in the literature \citep[e.g.,][]{Williamson2014,Wanderman2014,Clark2015}
and in this subsection we simply {\it summarize} their main conclusions.

\subsubsection{BNS mergers}
The sensitive distance can be approximated as  \citep{Clark2015}
\begin{equation}
D_{*,{\rm BNS}} \approx 400~{\rm Mpc}~(9/\rho_*),
\label{eq:D-1}
\end{equation}
where $\rho_*$ is the signal-to-noise of the GW signal.
When the advanced LIGO/VIRGO network has reached its full sensitivity,  with the `local' SGRB detection rate of
$4\pm 2 ~{\rm Gpc^{-3}~yr^{-1}}$ \citep{Wanderman2014},
the detection rate of the GW signal associated SGRB for a full sky $\gamma-$ray monitor is estimated as
\begin{equation}
{\cal R}_{\rm full~sky}(\rho_*=9) \approx 1\pm 0.5~{\rm yr^{-1}},
\label{eq:R-1}
\end{equation}
where all SGRBs are assumed to originate from BNS mergers. Such an assumption seems reasonable since the
BH$-$NS merger rate is generally expected to be just $\sim 1/10$
times that of the BNS merger rate \citep{Abadie2010}.
Please note that the LIGO/Virgo network can boost gravitational wave detection rates by exploiting the mass distribution of neutron stars within BNS system, and for searches with detected electromagnetic counterparts the detection rate may increase of 60\% \citep{Bartos2015}.

\subsubsection{BH-NS mergers}
No neutron star-black hole (NS-BH) binaries have not been directly observed yet \citep{Lattimer2012} but indirect evidence for
NS-BH merger has been suggested in the macronova modeling \citep{Yang2015,Jin2015}.
In a systematic analysis of the BH mass distribution based on 35 X-ray binaries, \citet{Farr2011} found strong evidence for a mass gap between the most massive neutron stars and the least massive black holes, confirming the results of \citet{Bailyn1998} and \citet{Ozel2010}. For the
low-mass systems (combined sample of systems), they found
a black hole mass distribution whose 1\% quantile lies above
4.3 $M_\odot$ (4.5 $M_\odot$) with 90\% confidence.  The typical NS-BH binary systems are expect to have a mass ratio of $\sim 1:4$, for which
the sensitive distance of the aLIGO/AdV network can be estimated as
\begin{equation}
D_{*,{\rm NS-BH}} \approx 690~{\rm Mpc}~(9/\rho_*).
\label{eq:D-2}
\end{equation}


If $\sim 1/5$ of short and long-short GRBs are produced by NS-BH mergers, while
aLIGO/AdV are more sensitive to the heavier NS-BH system, we expect around half
of the CBC events might have an EM counterpart has an origin of NS-BH mergers.\\

As an optimistic estimate (i.e., supposing most nearby SGRBs can be observed by Fermi GBM-like detectors), in 10 years of full run of aLIGO/AdV network $\sim 10-20$ GW-associated SGRBs are expected and possibly one half of them may have an NS-BH merger origin. The statistical study of such a sample, though still limited, may shed valuable light on the physical processes taking place at the central engine (see Sec. \ref{sec:lag} for the details) and possibly also the fundamental physics (see Sec. \ref{sec:speed} for the details). In addition to SGRBs, some supernova-less long GRBs may also have a compact object merger origin and the detection rate of merger events will increase.

\subsection{``Supernova-less" GRBs within the sensitivity distance of advanced LIGO/VIRGO network}
As shown in Sec. \ref{sec:SGRB-prospect}, the detection prospect of GW-associated SGRBs is promising for the advanced LIGO/VIRGO network.
{The nearest short burst is GRB 061201. It is measured to have a redshift of $z=0.111$ \citep[][see however D'Avanzo et al. (2014) for the uncertainty]{Berger2014}, or a distance of 520 Mpc, which is lager than $D_{*,{\rm BNS}}(\rho_*=9)$. Note that most SGRBs are expected to be powered by BNS mergers, {\it hence no single SGRB has been found within the averaged sensitive distance of the advanced LIGO/VIRGO network} \citep{Williamson2014,Wanderman2014}. Such a result is somewhat disappointing though not in significant tension with the expectation (see eq.(\ref{eq:R-1})).}
To better explore the situation, in this work we also take into account all ``nearby" (i.e., $z<0.3$) supernova-less long GRBs, including XRF 040701 \citep[X-ray flash,][]{Soderberg2005}, GRB 060505 and GRB 060614 \citep{Fynbo2006}. Please note that actually some ``SGRBs" with the so-called extended emission can also be classified as supernova-less long GRBs, but such events have also been included in previous GW/SGRB association studies. { Below we focus on the ``traditional"  long-short GRBs and introduce them in some details.}

XRF 040701 was localized by the Wide-Field X-Ray Monitor on board the {\it High Energy Transient Explorer (HETE-2)} on 2004 July 1.542 UT. It is characterized by the very low peak frequency (i.e., $<6$ keV) of the prompt emission. Soderberg et al. (2005)'s  foreground extinction-corrected HST detection limit is $\simeq 6$ mag fainter than SN 1998bw, the archetypal hypernova that accompanied long GRBs \citep{Galama1998},
at a redshift of $z=0.21$. The analysis of the X-ray afterglow spectra reveals that the rest-frame host galaxy extinction is constrained
to $A_{\rm V,host}<2.8$ mag, suggesting that the associated supernova, if there was,
should be at least $\sim $3.2 mag fainter than SN 1998bw  \citep{Soderberg2005}.
Due to the lack of the sufficient
multi-wavelength afterglow data, the ``absence"
of a bright supernova associated with XRF 040701 did not attract wide attention.
The situation changed dramatically when the supernovae associated with GRB 060505
and GRB 060614 had not been detected down to limits hundreds of times fainter than
SN 1998bw \citep{Fynbo2006}. Particularly, GRB 060614, a bright burst with a duration
of $\sim 102$ s at a redshift of $0.125$, had dense followup observations with Very
Large Telescope and Hubble Space Telescope. The physical origin (either a peculiar
collapsar or a compact object merger) of GRB 060614 was debated over years
\citep[e.g.,][]{Fynbo2006,Gehrels2006,DellaValle2006,Gal-Yam2006,Zhang2007}.
The re-analysis of the optical afterglow emission of GRB 060614 found
significant excess components in multi-wavelength photometric
observations, which can be reasonably interpreted as a Li-Paczy\'{n}ski
 macronova that powered by the radioactive decay of debris
following an NS-BH merger while the weak supernova model does not work
\citep{Yang2015,Jin2015}. As summarized in \citet{Xu2009}, the origin of
GRB 060505 at a redshift of $z=0.089$ is less clear. The properties of
its host galaxy seems to be consistent with that expected for the
long-duration GRBs \citep{Thone2008} but GRB 060505 is an outlier of the
so-called Amati relation that holds for long GRBs \citep{Amati2007}. The
HST observations at $t\sim 14.4$ days after the burst did not find optical
 emission down to a limiting AB magnitudes of 27.3$^{\rm th}$ in F814W band
 and 27.1$^{\rm th}$ in F475W band \citep{Ofek2007}. Such stringent limits
 are strongly at odds with the collapsar model but can be well consistent
 with the BNS merger model as long as the r-process ejecta has a mass $<10^{-3}~M_\odot$ \citep{Jin2016}.

\begin{figure}
\begin{center}
\includegraphics[width=0.5\textwidth]{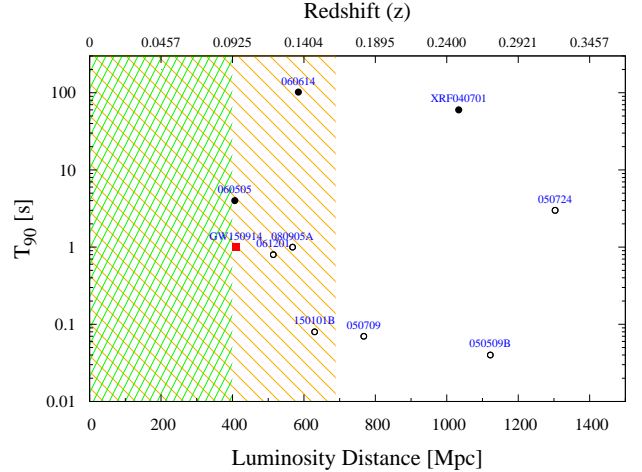}
\end{center}
\caption{The ``nearby" (i.e., $z\leq 0.3$) {\it supernova-less} GRBs, including the SGRBs and the long-short GRBs, and the averaged sensitivity distances of advanced LIGO/VIRGO network (i.e., $\rho_*\geq 9$) for binary neutron star mergers and neutron star-stellar black hole mergers. The open circles are the SGRBs discussed in the literature, including GRB 050724 with a duration $T_{90}=3\pm 1$ s \citep{Fox2005}. The red filled circles are the long-short GRBs. The data are taken from \citet{Berger2014}, \citet{Fynbo2006}, \citet{Soderberg2005} and \citet{Levan2015}. Interestingly, the long-short burst GRB 060505 is within $D_{*, \rm BNS}(\rho_*=9)$. Three short bursts (including GRB 061211, GRB 080905A and GRB 150101B) and the long-short burst GRB 060614 are within $D_{*, \rm NS-BH}(\rho_*=9)$. Though the NS-BH merger rate is expected to be one order of magnitude lower than that of the BNS mergers and hence most supernova-less GRBs should have a BNS merger origin, there is some evidence for an NS-BH merger origin of GRB 060614. Hence the GW signals of GRB 060614-like events taking place in the era of advanced LIGO/VIRGO would be detectable. The adopted cosmological model parameters is presented in detail in Sec.\ref{sec:speed-1}. {GW150914 is also included in the plot for illustration. We remind the readers that the observed gravitational wave event is a binary black hole coalesce. Although for O1, advanced LIGO detectors are not yet sensitive enough to detect NS-BH events from that distance, this observation of GW150914 demonstrates that CBC events do happen within the NS-BH horizon of the full sensitive advanced LIGO/Virgo network.}
}
\label{fig:1}
\end{figure}

 It may be still a bit early to conclude that all ``nearby supernova-less" long GRBs (i.e., ``long-short GRBs") are from compact object binary mergers. The successful identification of a macronova signal in the long-short event GRB 060614, nevertheless, renders such a possibility more attractive than before. If the NS-BH merger model for GRB 060614 is correct, the luminosity distance of this event is $D_{\rm L}\approx 576$ Mpc, which is smaller than $D_{\rm *,NS-BH}$ as long as $\rho_*\leq 10.8$ (see eq.(\ref{eq:D-2})). For GRB 060505, the redshift $z=0.089$ corresponds to a luminosity distance $D_{\rm L}\approx 400$ Mpc, which almost equals to $D_{\rm *,BNS}$ for $\rho_*=9$ (see eq.(\ref{eq:D-1})). {\it Intriguingly, among the supernova-less and short events detected so far (note that the GBM transient 150914 is still uncertain), the long-short burst GRB 060505 and GRB 060614 are the only candidates that might yield detectable GW signal for the advanced LIGO/VIRGO network} (see Fig.\ref{fig:1}). .

The presence of two {\it Swift} GRB candidates within the averaged sensitivity distance of the advanced LIGO/VIRGO network for $\rho_*\geq 9$ is indeed very encouraging for the ongoing GW experiments. On the other hand, if the supernova-less long-duration XRFs/GRBs were from a peculiar kind of collapsar (which is very unlikely to be the case for GRB 060614), we can verify it with the non-detection of GW signal. Therefore we suggest that the supernova-less long-duration XRFs/GRBs are one of prime targets for advanced LIGO/VIRGO network and the nature of such a kind of ``mysterious" events would be unambiguously pinned down in the era of GW astronomy.

\section{Model-dependent estimates of time lag between binary coalescence and GRB onset}\label{sec:lag}
{Though extensively studied, so far the launching, acceleration and energy dissipation of the GRB ejecta are still heavily debated \citep[see][for a dedicated review]{Zhang2014}. Instead of carrying out advanced study of a specific model, in this section we adopt some widely-discussed scenarios} and present our model-dependent estimates of time lag between GW coalescence time and GRB onset (i.e., $\Delta t_{\rm GW-GRB}$) and {examine} how a statistical study of $\Delta t_{\rm GW-GRB}$ could help us to better understand the physical processes taking place at the central engine.

In general, $\Delta t_{\rm GW-GRB}$
 can be divided into two parts. One is the time delay between the
merger time, which could be estimated by analyzing the GW data \citep[e.g.,][]{Fairhurst2011,Veitch2014},
and the successful launch of the ultra-relativistic ejecta (i.e., $\Delta t_{\rm laun}$).
The other is the time delay between the  launch of the ultra-relativistic ejecta and the onset of
the gamma-ray emission (i.e., $\Delta t_{\rm em}$). Below we examine $\Delta t_{\rm laun}$ and $\Delta t_{\rm em}$
separately under different models and then get the corresponding estimate of
\[\Delta t_{\rm GW-GRB}=\Delta t_{\rm laun}+\Delta t_{\rm em}.\]
Notice that we are using the coalescence time, when the GW signal spikes, as a proxy to the merger moment.
One might reasonably argue that these two are not identical, but as the binary system evolve very rapidly towards the merger,
they should not differ by more than a couple of rotations around the merger.
For BNS systems the merger frequency is around $1000$Hz so the difference in time
can not exceed $\sim 1$ms \citep{Fairhurst2011,Purrer2014}.

{In this work we focus on the most widely-adopted hypothesis that the SGRBs were powered quickly after the merger (i.e., $\Delta t_{\rm GW-GRB}<10$ s). However notice that a small amount of SGRBs seem to have precursor emission and the precursors are likely from the same central engine activity as the main bursts \citep{Charisi2015}. In the binary merger scenario, the merger
may likely happened before the precursor. If so, the time-lag between the GW signal and the SGRB/long-short GRB can be long to $\sim 100$ s, which may label the lifetime of the supramassive neutron star formed in the merger or alternatively the fall-back accretion timescale of the fragmented part of the compact object. \citet{Rezzolla2015} argued that the BNS mergers might actually take place several hours before the SGRBs. The GW/GRB association observations can easily distinguish between such scenarios (i.e., $\Delta t_{\rm GW-GRB}\sim 10^{2}-10^{4}$ s) with and the short delay cases (i.e., $\Delta t_{\rm GW-GRB}<10$ s), as shown in Tab. \ref{tab:Summary}.  A long time-lag between the GW and GRB signals, in principle, could also arise from the supluminal movement of the GW in the vacuum or its higher velocity than the photons in the gravitational potential. Such possibility can be precisely tested as long as a sample of GW/GRB association is established (see Sec.\ref{sec:EEP} for the details).}

\subsection{$\Delta t_{\rm laun}$ expected in different merger scenarios}
\subsubsection{NS-NS mergers}\label{sec:NS-NS-mergers}
The maximum gravitational mass of a cold non-rotating neutron star is known to be $M_{\rm max}>2~M_\odot$ \citep{Antoniadis2013} and
the threshold for collapse of the merger-formed remnants into black holes can be estimated roughly as $M_{\rm thres}\approx 1.35~M_{\rm max}$ \citep{Shibata2006}
Hence a total gravitational mass $\gtrsim2.7~M_\odot$ is likely required for prompt collapse to a black hole. Such massive neutron star binaries should just account for a small fraction of merger events if the mass distribution of ``cosmic" neutron star binaries resembles what was observed in the Galaxy \citep[see][for a recent review]{Lattimer2012}. Hence usually we do not expect the prompt collapse of the merger-formed neutron stars. Instead the merger formed remnants are widely expected to be very massive neutron stars with strong differential rotation that can support against the collapse at least temporarily.

Such remnants are called the hypermassive neutron stars (HMNSs).
The fate of the post-merger HMNSs is however uncertain, and is contingent on the mass limit for
support of a hot, differentially rotating configuration \citep[e.g.,][]{Baumgarte2000,Hotokezaka2013}.
Below we present the model-dependent estimates of the collapse time of the HMNSs which we regard as $\Delta t_{\rm laun}$.
{One exception would be presented in the last paragraph of this sub-subsection.}

When $M_{\rm max}<M<M_{\rm thres}$,
various mechanisms could act to dissipate and/or transport
energy and angular momentum, possibly inducing collapse after
a delay which could range from tens of milliseconds to a few
seconds \citep[see][for a recent review]{Faber2013}. For instance, in the presence of strong magnetic field, the magnetic braking effect can effectively transfer the angular-momentum in a timescale
$\tau_{\rm br}\sim {R_{\rm s}/V_{\rm A}}\sim 0.3~{\rm s}~({R_{\rm s}/10^{6}~{\rm cm}})({\rho/ 10^{15}~{\rm g~cm^{-3}}})^{1\over 2}({\epsilon / 0.3})({B_{\rm s}/10^{13}~{\rm G}})^{-1}$,
where $V_{\rm A}$ is the Alfven's velocity, $R_{\rm s}$ is the radius of the neutron star and $\epsilon \sim 0.3$ is the expected strength ratio between the surface magnetic field $B_{\rm s}$ and the interior polorial magnetic field  \citep{Shapiro2000}.  Another mechanism is the magnetorotational instability (MRI), which generates turbulence in a magnetized rotating fluid body that amplifies the magnetic field and transfers angular momentum. In the presence of MRI, an effective viscosity is likely to be generated with the effective viscous parameter
$\nu_{\rm vis}\sim \alpha_{\rm vis} c_{\rm s}^{2}/\Omega_{\rm c}$,
where $\alpha_{\rm vis}$ is the viscosity parameter, $c_{\rm s}$ is the sound velocity of the envelope of the HMNS and $\Omega_{\rm c}$ is the angular velocity of the core of the differentially-rotating neutron star \citep{Balbus1991}.
Thus, the viscous angular momentum transport time scale can be estimated as
$\tau_{_{\rm MRI}}\sim R_{\rm s}^2/\nu_{\rm vis} \sim 0.1~{\rm s}~({R_{\rm s}/10^{6}~{\rm cm}})^2({\alpha_{\rm vis}/0.01})^{-1}({c_{\rm s}/ 0.1c})^{-2}({\Omega_{\rm c}/10^4~{\rm rad~s^{-1}}})^{-1}$ \citep{Hotokezaka2013,Siegel2013}. A reasonable estimate of the termination timescale of the differential rotation is $\tau_{\rm diff}=\min\{\tau_{\rm br}, \tau_{_{\rm MRI}}\}\sim 0.1$ s, after which the HMNS is expected to collapse.

The situation is even less uncertain when both finite-temperature
effects in the equation of state and neutrino emission of the central compact object have been taken into account. In the numerical simulations of the merger of binary neutron stars performed in full general relativity incorporating the finite-temperature effect and neutrino cooling, \citet{Sekiguchi2011} found that the effect of the thermal energy is significant and can increase $M_{\rm max}$ by a factor of $20\%-30\%$ for a high-temperature state with $T\geq 20$ MeV. Since they are not supported by differential rotation, the hypermassive remnants were predicted to be stable until neutrino cooling, with luminosity of $\sim 3-10\times 10^{53}~{\rm erg~s^{-1}}$, has removed the pressure support in $\tau_{\rm thermal}\sim 1~{\rm s}$ \citep{Sekiguchi2011}.

For $t<\tau_{\rm w}$, a baryon-loaded wind is continuously ejected which would bound the bulk Lorentz factor of the jet
to $\Gamma_{\rm w}\sim5(L_{\rm jet}/10^{52}{\rm erg~s^{-1}})(\dot{M}_{\rm w}/10^{-3}M_\odot~{\rm s}^{-1})^{-1}$, where $L_{\rm jet}$ is the isotropic-equivalent luminosity of the jet, $\dot{M}_{\rm w}$ is the mass loss rate via the wind and $\tau_{\rm w}$ is the wind duration (either $\tau_{\rm diff}$ or $\tau_{\rm thermal}$, depending on the mechanism that mainly supports against the collapse). Such a low $\Gamma_{\rm w}$ is too small to give rise to energetic GRB emission. Hence it is widely anticipated that no GRB is possible unless the neutrino-driven wind gets very weak or more realistically the neutron star has collapsed to a black hole. After the collapse of the HMNS, the earlier out-moving dense wind remains to hamper the advance of the
jet, whose injection lifetime, $t_{\rm jet}$, is determined by the viscous
timescale of the accretion disk (Lee et al. 2004). \citet{Murguia-Berthier2014} suggested that in the black hole central engine model for (short) GRBs, the black hole formation should occur promptly as any moderate delay at the hyper-massive neutron star stage
would result in a choked jet. The argument that just the mergers with a remnant collapse within a timescale $\sim t_{\rm diff}\sim 0.1$ s can produce (short) GRBs may just hold in the scenario of energy extraction via neutrino mechanisms. The accretion timescale of the torus formed in binary neutron star mergers can be estimated as $t_{\rm acc}\sim 0.1~{\rm s}~(\alpha_{\rm vis}/0.1)^{-6/5}$ \citep{Narayan2001,Popham1999}. Following \citet{Zalamea2011} and \citet{Fan2011}, it is straightforward to estimate the corresponding luminosity of the annihilated neutrinos/antineutrinos as $L_{\nu\bar{\nu}}\approx 10^{49}~{\rm erg~s^{-1}}(\dot{m}/0.1~M_\odot~{\rm s}^{-1})^{9/4}$, where $\dot{m}$ is the accretion rate and the spin of the black hole has been taken to be $a=0.78$, a typical value for the black hole formed in binary neutron star mergers. The isotropic-equivalent luminosity of the ejecta is then $L_{\rm jet}\approx 2\times 10^{51}~{\rm erg~s^{-1}}(\dot{m}/0.1~M_\odot~{\rm s}^{-1})^{9/4}(\theta_{\rm jet}/0.1)^{-2}$, which satisfies the condition of relativistic expansion of the jet head within the preceding neutrino-driven wind medium (i.e., eq.(8) of \citet{Murguia-Berthier2014}) as long as $\dot{m}>\dot{m}_{\rm jet}\approx 0.2~M_\odot~{\rm s^{-1}}$. The accretion disk mass is $M_{\rm disk}\sim \dot{m}_{\rm jet}t_{\rm acc}\sim 0.02~M_\odot$. Such an accretion disk mass is consistent with that found in numerical simulations of binary neutron star mergers \citep{Faber2013,Nagakura2014}, which in turn suggests that short GRBs are possible for $t_{\rm jet}\approx t_{\rm acc}>\tau_{\rm w}$ if $\tau_{\rm w}<0.1$ s, in agreement with \citet{Murguia-Berthier2014}. If instead $\tau_{\rm w}\gg 0.1$ s, the required $M_{\rm disk}$ would be too massive to be realistic \citep{Fan2011,Liu2015}.

The situation is significantly different for the magnetic process to launch the GRB ejecta. The huge amount of rotational energy of the black hole can be extracted efficiently via the Blandford$-$Znajek process and the luminosity of the electromagnetic outflow can be estimated by $L_{\rm BZ}\approx 6\times10^{49}~{\rm erg~s^{-1}}~(a/0.75)^{2}(B_{\rm H}/10^{15}~{\rm G})^{2}$, where $B_{\rm H} \sim 1.1\times 10^{15}~{\rm G}~(\dot{m}/0.01~M_\odot~{\rm s}^{-1})^{1/2}(R_{\rm H}/10^{6}~{\rm cm})^{-1}$ is the magnetic field strength on the horizon of the black hole \citep{Blandford1977}. Therefore $\dot{m}\sim 0.01~M_\odot~{\rm s}^{-1}$ is sufficient to launch energetic ejecta with $L_{\rm jet} \sim 10^{52}~{\rm erg~s^{-1}}(\theta_{\rm j}/0.1)^{-2}$. An $\alpha\leq 0.01$ is needed to get a $t_{\rm acc} \sim$ a few seconds. Such a ``small" $\alpha$ is still possible \citep{Narayan2001} and the required accretion disk mass is also in the reasonable range of $\sim 0.01~M_\odot$. Please note that in these estimate the ejecta ``breakout" criterion suggested in \citet{Murguia-Berthier2014} has been adopted. In reality, the Poynting-flux jet could break out from the ``neutrino-driven wind" more easily than the hydrodynamic jet. This is because the reverse shock that slows down the hydrodynamic jet and the collimation shock that collimates it, cannot form within
the Poynting-flux-dominated jet. As a result the Poynting-flux dominated jet moves much faster and dissipates much
less energy while it crosses the preceding neutrino-driven wind \citep{Bromberg2014}. The latest time-dependent 3D relativistic magnetohydrodynamic simulations of relativistic, Poynting-flux dominated jets that propagate into medium with a spherically-symmetric power-law density
distribution has found out that some instabilities can  leads to efficient
dissipation of the toroidal magnetic field component and hence the propagation of such a ``headed" magnetized ejecta is likely similar to that of a hydrodynamic ejecta \citep{Bromberg2015}. In such a case, the ``breakout" criterion of \citet{Murguia-Berthier2014} applies. After the ``breakout" of the ``headed" magnetized ejecta, an evacuated
funnel presents and the later ejecta moves freely without significant magnetic energy dissipation (i.e., it is within the phase of ``headless" jet \citep{Bromberg2015}). Therefore, a $t_{\rm acc}\sim $ a few seconds may be sufficiently long to successfully produce GRBs for $\tau_{\rm w}\sim \tau_{\rm thermal}\sim 1$ s. The conclusion of this paragraph is that GRB is still possible in the case of $\tau_{\rm w}\sim \tau_{\rm thermal}\sim 1$ s but the outflow should be launched via magnetic processes.

The expected time delay between the merger of the binary neutron stars and the launch of the ultra-relativistic GRB outflow can thus be approximately summarized as $\Delta t_{\rm laun,BNS}\sim (0.01~{\rm s},~0.1~{\rm s},~1~{\rm s})$ for (the prompt formation of black hole, differential rotation supported HMNS, thermal pressure supported HMNS), respectively. The minimum $\Delta t_{\rm laun,BNS}$ is taken to be $\sim 10$ ms since the  merger time is expected to be measured with an accuracy better than $\sim 10$ ms and the ultra-relativistic outflow may be launched
promptly.

{In the above discussion we assume that the short GRBs are produced when the HMNSs collapse into black holes. There is another possibility that the differentially-rotating NSs can eject a significant material towards the rotation axis which might also produce (short) GRBs. Such a scenario has attracted wide attention since the analysis of a good fraction of afterglow emission of  {\it Swift} short GRBs found possible evidence for the magnetar central engine \citep{Rowlinson2013}. One possible physical scenario is that the differentially rotating neutron star wraps the poloidal seed magnetic field into super-strong toroidal fields ($B_{\rm f}\sim 10^{17}$ Gauss) that may emerge from the star through buoyancy and then generate GRBs via magnetic energy dissipation \citep{Kluzaniak1998,Dai2006}. In this model $\Delta t_{\rm laun, BNS}$ is expected to be the time needed to amplify the seed magnetic field to $B_{\rm f}\sim 10^{17}$ Gauss, i.e., $\Delta t_{\rm laun, BNS} \sim 5~{\rm ms}~({B_{\rm f}/ 10^{17}~{\rm G}})({\epsilon/0.3})({B_{\rm s}/10^{15}~{\rm G}})^{-1}({\Delta \Omega / 6000~{\rm rad~s^{-1}}})^{-1}$, where $\Delta \Omega \equiv 2\pi (1/P_{\rm c}-1/P_{\rm s})$, and $P_{\rm c}$ and $P_{\rm s}$ are the rotational periods of the differentially rotating internal part and main NS, respectively \citep{Kluzaniak1998}. With the $B_{\rm s}$ and the initial rotational periods of the magnetar central engine estimated in \citet{Rowlinson2013} we have $\Delta t_{\rm laun, BNS} \sim 10-100~{\rm ms}$. }

\subsubsection{NS-BH mergers}
In this case the central engine is a stellar mass black hole and the region along the spin axis
of the black hole is likely cleaner than the case of NS-NS mergers. However, the joint effects of shocks during the
disk circularization, instabilities at the disk/tail interface, and
neutrino absorption unbinds a small amount ($\sim 10^{-4}~M_\odot$) of material in the
polar regions \citep{Foucart2015}. Over longer timescales, the neutrino-powered
winds become active and eject material in the polar regions. Though the material  is still negligible
compared to the material ejected dynamically in the equatorial
plane during the disruption of the neutron star, this ejecta could impact the formation of a relativistic jet
\citep{Foucart2015}. Nevertheless, a few percents of the energy radiated in neutrinos
is expected to be deposited in the region along the spin axis
of the black hole through $\nu\bar{\nu}$ annihilations \citep{Setiawan2006,Janka1998}.
The energy deposition at a rate $\sim 10^{51}~{\rm erg~s^{-1}}$ might also be able to
power short $\gamma$-ray burst \citep{Foucart2014,Lee2007}.

As in the BNS merger scenario, the magnetic mechanism may be more promising in launching ultra-relativistic outflows and then giving rise to GRBs.   In the recent high-resolution numerical-relativity simulations for the merger of BH-NS
binaries  that are subject to tidal disruption and subsequent formation of a massive accretion torus, the accretion torus formed quickly and the magnetic-field was amplified significantly due to the non-
axisymmetric magnetorotational instability and magnetic
winding \citep{Paschalidis2015,Kiuchi2015}. The amplification can yield $B\sim 10^{15}$ G
at the BH poles in $\sim 20$ milliseconds after the merger and the corresponding Blandford-Znajek luminosity can be sufficient high to power GRBs.

For the role of the magnetic process in extracting the energy for the GRBs, the data of GRB 060614 likely has shed valuable light on. Such a  long-short event is most likely powered by the merger of a binary system of neutron star and stellar mass black hole \citep{Yang2015,Jin2015}. As found in various numerical simulations, the total mass of the accretion disk is expected to be not much more massive than $\sim 0.1~M_\odot$. On the other hand, the duration of the ``long-lasting" soft $\gamma-$ray emission is $\sim 100$ s. Hence the time averaged accretion rate is expected to be just in order of $\dot{m}\sim 10^{-3}~M_{\odot}~{\rm s}^{-1}$. For such a low accretion rate, the neutrino mechanism is expected to be unable to launch energetic GRB outflow \citep[e.g.,][]{Fan2005,Liu2015}. Instead, the Blandford$-$Znajek process can give rise to Poynting-flux dominated outflow with an ``intrinsic" luminosity $L_{\rm BZ}\approx 6\times10^{47}~{\rm erg~s^{-1}}~(a/0.75)^{2}(B_{\rm H}/10^{14}~{\rm G})^{2}$ \citep{Blandford1977}, which is sufficient to explain the observed $\gamma-$ray luminosity of GRB 060614 after the correction of a jet opening angle of the outflow $\theta_{\rm j}\sim 0.1$ \citep[see][]{Xu2009}. Therefore, the soft long-lasting gamma-ray ``tail" emission of GRB 060614 likely has a moderate to high linear polarization \citep{Fan2005}.

In view of these facts, we suggest that ultra-relativistic outflows may be launched within $\Delta t_{\rm laun,BHNS}\sim$10 milliseconds after the BH-NS mergers via either the neutrino-antineutrino annihilation or magnetic process(es). \\

\subsection{$\Delta t_{\rm em}$ expected in baryonic and magnetic outflow models}
\subsubsection{The bayronic outflow}
The neutrino-antineutrino annihilation process will launch an extremely-hot fireball. For such a kind of baryonic outflow, the acceleration is well understood \citep{Piran1993,Meszaros1993} and most of the initial thermal energy
may have been converted into the kinetic energy of the
baryons at the end of the acceleration \citep{Shemi1990}.
A quasi-thermal emission component, however, is likely inevitable \citep[see][and the references therein for the resulting spectrum]{
Chhotray2015}. The quasi-thermal emission is mainly from
the photosphere at a radius $R_{\rm ph}$, which can be estimated as
$R_{\rm ph}\approx 4.6\times 10^{10}~{\rm cm}~(L/10^{51}~{\rm erg~s^{-1}})(\eta/200)^{-3}$,
where $L$ is the total isotropic-equivalent luminosity of the baryonic outflow and {$\eta\sim 10^{2}-10^{3}$ is the initial dimensionless entropy} \citep{Paczynski1990,Daigne2002}. Assuming an initial launch radius of the fireball $R_0\approx 10^{7}$ cm, at $R_{\rm ph}$ the thermal radiation luminosity is expected to be $L_{\rm th}\approx 2.5\times 10^{49}~{\rm erg~s^{-1}}(L/10^{51}~{\rm erg~s^{-1}})^{1/3}(\eta/200)^{8/3}(R_0/10^{7}~{\rm cm})^{2/3}$. And the quasi-thermal radiation peaks at a temperature $T_{\rm th}\sim 80~{\rm keV}~(L/10^{51}~{\rm erg~s^{-1}})^{1/4}(\eta/200)^{2/3}(R_0/10^{7}~{\rm cm})^{1/6}$. For the sources within the advanced LIGO/VIRGO detection ranges (i.e., $D\approx 300$ Mpc), the energy flux can be high up to ${\cal F}\sim 2\times 10^{-6}~{\rm erg~s^{-1}}~(L/10^{51}~{\rm erg~s^{-1}})^{1/3}(\eta/200)^{8/3}(R_0/10^{7}~{\rm cm})^{2/3}$, which is detectable for {\it Swift} or {\it Fermi}-GBM. The acceleration timescale of the baryonic is $\sim (1+z)R_0/c$ and the delay between the termination of the acceleration and the emergence of the thermal photons can be estimated as $\sim (1+z)R_{\rm ph}/2\eta^{2}c$. For $\eta\geq 100$ the latter is significantly smaller than the former, hence $\Delta t_{\rm em}\sim (1+z)R_0/c \sim 0.3~{\rm ms}~(1+z)(R_{\rm 0}/10^{7}~{\rm cm})$, which is ignorably small.

If the photospheric quasi-thermal radiation is non-detectable (say, for the NS-BH mergers at $D \sim 1$ Gpc), more efficient emission may be cased by the collision between the baryonic shells ejected from the same central engine but with much different Lorentz factors. Strong internal shocks are generated and ultra-relativistic particles are accelerated. A fraction of internal shock energy has been converted into magnetic field and the electrons moving in the magnetic field produce energetic $\gamma-$ray emission \citep{Rees1994}. In such a model, the variability of the prompt emission largely traces the behavior of the activity of the central engine and the onset of the ``internal shock emission" is expected to be within the typical variability timescale of the prompt emission that can be $\sim 1-10$ ms \citep{Piran1999}, i.e., $\Delta t_{\rm em}
\sim 1-10$ ms.

\subsubsection{The magnetic outflow}
In the case of the magnetic outflow, both the acceleration and the subsequent energy-dissipation/radiation are more uncertain \citep[see][for recent reviews]{Kumar2015,Granot2015}. If the magnetic energy has been effectively converted into the kinetic energy of the outflow \citep{Granot2015}, the prompt emission of GRBs can be from the magnetized internal shocks \citep{Fan2004} or the photosphere with internal dissipation of energy via gradual magnetic reconnection \citep{Giannios2008} and we expect $\Delta t_{\rm em} \sim 1-10$ ms. Note that in the latest time-dependent 3D relativistic magnetohydrodynamic simulations of relativistic, Poynting-flux dominated jets that propagate into medium with a spherically-symmetric power-law density distribution has found out that some instabilities can  leads to efficient
dissipation of the toroidal magnetic field component \citep{Bromberg2015}, for which the onset of the prompt emission is likely dominated by the photospheric radiation and $\Delta t_{\rm em}$ is ignorably small.

If the photospheric radiation is too weak to be detectable for {\it Fermi}-GBM-like detectors due to either the absence of a dense wind-like medium in the direction of the black hole spin or the small luminosity of the breaking out material, the observed onset of the prompt emission is likely significantly delayed. In some models most of the initial magnetic energy has not been converted into the kinetic/thermal energy of the outflow and the prompt emission of GRBs is due to the magnetic energy dissipation at a rather large distance $R_{\rm pro}\sim 10^{16}$ cm \footnote{The magnetized internal shocks with significant magnetic dissipation \citep{Fan2004} can take place at a much smaller radius, say, $\sim 10^{14}-10^{15}$ cm.} possibly due to the breakdown of magnetohydrodynamic approximation of the highly-magnetized outflow \citep{Usov1994,Zhang2002,Fan2005}, or the current-driven instabilities developed in the outflow shell \citep{Lyutikov2003}, or the internal collision-induced magnetic reconnection and turbulence \citep{Zhang2011}. {Correspondingly we have $\Delta t_{\rm em} \approx (1+z)R_{\rm pro}/2\eta^2c \approx 2~{\rm s}(1+z)(R_{\rm pro}/10^{16}~{\rm cm})(\eta/300)^{-2}$ (i.e., the radial timescale), which equals to the ``angular timescale" of the emission, the minimal timescale of the GRB duration, as long as the ejecta has an opening angle larger than $1/\eta$ \citep{Piran1999}. Since the angular timescale is derived for an infinite-thin radiating shell, it is expected to be shorter than the duration of the prompt emission of the whole GRB (i.e., $\Delta t_{\rm em}\leq T_{90}$).}

SGRB 050509B and SGRB 050709 have $T_{90}=0.04$ s and $0.07$ s, respectively \citep{Fox2005}. For such ``brief" events, $R_{\rm pro}\sim 10^{16}$ cm is {\it disfavored} unless $\eta \gtrsim 2000$. The $\eta$ as high as $\sim 2000$, however, would render them the outstanding outliers of the correlation $\eta\approx 250~(L_\gamma/10^{52}~{\rm erg~s^{-1}})^{0.3}$ holding for some long GRBs \citep{Lv2012,Fan2012,Liang2015} and possibly also the short burst GRB 090510 if its $\eta \gtrsim 1200$, as argued in \citet{Ackermann2010}, where $L_\gamma$ is the $\gamma-$ray luminosity of the GRB. Moreover, unless there is the fine tuning that the central engine shut down almost at the same time as the outflow breaks out the dense material, the central engines of these two very-short bursts should be (promptly-formed) black holes and $\Delta t_{\rm GW-GRB}<T_{90}$ is expected. For SGRB 050709 the modeling of the macronova signal favors a NS-BH merger origin \citep{Jin2016}, which is in support of our current argument (i.e., the central engine of SGRB 050709 was a black hole).

\subsection{Expected relationship between $\Delta t_{\rm GW-GRB}$ and $T_{90}$}
We summarize in Table 1 the suspected $\Delta t_{\rm GW-GRB}$ (i.e., the sum of $\Delta t_{\rm laun}$ and $\Delta t_{\rm em}$), where the case of $R_{\rm pro}\ll 10^{16}$ cm includes the scenarios of photospheric radiation and regular (magnetized) internal shock radiation. Clearly, the shortest delay is expected in the cases of prompt BH formation in the NS-NS mergers or the NS-BH mergers if the onset of the prompt emission is governed by the photosphere or regular internal shocks (i.e., $R_{\rm pro}\ll 10^{16}$ cm) and such events will be valuable in imposing very stringent constraint on the difference between the GW and the speed of light (see Sec.\ref{sec:speed} for the details).

In Sec.\ref{sec:NS-NS-mergers} we have already mentioned that the prompt formation of BH in BNS mergers is likely uncommon. As long as $R_{\rm pro}\ll 10^{16}$ cm, one naturally expects that (1) for BNS mergers $\Delta t_{\rm GW-GRB}$ is significantly longer than that of the NS-BH mergers, i.e., $\Delta t_{\rm GW-GRB}({\rm NS-BH})\ll \Delta t_{\rm GW-GRB}({\rm BNS})$; (2) for NS-BH mergers, usually $\Delta t_{\rm GW-GRB}$ is expected to be shorter than $T_{90}$, i.e., $\Delta t_{\rm GW-GRB}({\rm NS-BH})<T_{90}$. While for $R_{\rm pro}\sim 10^{16}~{\rm cm}$, we expect that $\Delta t_{\rm GW-GRB}$ should be comparable with $T_{90}$ for both BNS and NS-BH merger powered SGRBs (Note that for some very-shortly lasting events such as SGRB 050509B and SGRB 050709, $R_{\rm pro}\sim 10^{16}$ cm is most-likely disfavored). Therefore, with reasonable large BNS merger GRB sample and NS-BH merger GRB sample, the statistical distribution of $\Delta t_{\rm GW-GRB}$ and $T_{90}$ for each sample or alternatively the distribution of $\Delta t_{\rm GW-GRB}$ for the combined sample could shed valuable light on the central engine physics.

\begin{table*}
\caption {Expected time delay between the coalescence and the GRB onset (i.e., $\Delta t_{\rm GW-GRB}$)$^{a}$.}
\begin{tabular}{llll}
\hline
Mergers & Prompt Remnant & $R_{\rm pro}\ll 10^{16}$ cm & $R_{\rm pro}\sim 10^{16}$ cm$^{b}$ \\
\hline
          & BH         & $\sim 10$ ms  &  $\sim2~{\rm s}(1+z)({R_{\rm pro}/10^{16}~{\rm cm}})({\eta/300})^{-2}$\\
  BNS     & DRS$^{b}$ HMNS   & $\sim 100$ ms &  $\sim 0.1~{\rm s}+2~{\rm s}(1+z)({R_{\rm pro}/10^{16}~{\rm cm}})({\eta/300})^{-2}$\\
          & TPS$^{b}$ HMNS   & $\sim 1$ s    &  $\sim 1~{\rm s}+2~{\rm s}(1+z)({R_{\rm pro}/10^{16}~{\rm cm}})({\eta/300})^{-2}$\\
          & GRB$-$DiffNS$^{c}$   & $\sim 10-100$ ms    &  $\sim 0.1~{\rm s}+2~{\rm s}(1+z)({R_{\rm pro}/10^{16}~{\rm cm}})({\eta/300})^{-2}$\\
\hline
NS$-$BH   & BH         & $\sim 10$ ms  &  $\sim 2~{\rm s}(1+z)({R_{\rm pro}/10^{16}~{\rm cm}})({\eta/300})^{-2}$\\
\hline
\end{tabular}
\label{tab:Summary}\\
\begin{minipage}{18cm}
$^{a}$ {Note that in some specific cases $\Delta t_{\rm GW-GRB} \sim 10^{2}-10^{4}$ s are possible \citep[e.g.][]{Charisi2015,Rezzolla2015}, which can be easily distinguished from the scenarios summarized in this table as long as a sample of GW/GRB association has been established.}\\
$^{b}$ Note that in general $T_{90}\geq 2~{\rm s}(1+z)(R_{\rm pro}/10^{16}~{\rm cm})({\eta/300})^{-2}$. For SGRs, unless $(R_{\rm pro}/10^{16}~{\rm cm})({\eta/300})^{-2}\ll 1$, $\Delta t_{\rm GW-GRB}$ is expected to be comparable with $T_{90}$.\\
$^{c}$ DRS is the abbreviation of ``Differential rotation supported" and TPS is the abbreviation of ``Thermal pressure supported".\\
{$^{d}$ GRB$-$DiffNS represents the case of that differentially-rotating NS directly launches GRB outflow (see the last paragraph of Sec.\ref{sec:NS-NS-mergers} for the discussion on such a kind of possibility). This case is different from the first three scenarios in which the GRB ejecta is assumed to be launched when the gravitational collapse takes place.}
\end{minipage}
\end{table*}

Are BNS mergers and NS-BH merger events distinguishable in the era of advanced LIGO/VIRGO? It is known that GW observations can efficiently measure the binary's chirp mass ${\cal M}\equiv(m_1m_2)^{3/5}/(m_1+m_2)^{1/5}$, which however leaves the individual masses undetermined, where $m_1$ and $m_2$ are the gravitational masses of the binary stars, respectively \citep[e.g.][]{Bartos2013,Hannam2013}. Moreover, the accuracy of the reconstruction of the
masses is decreased by the additional
mass-ratio-spin degeneracy. Fortunately, in many cases the nature of the binary system can be determined. For instance, considering non-spinning compact objects and a $\rho_*\approx 10$, a ${\cal M} \gtrsim 2.8 M_\odot$ implies that one of the binary
compact objects has to have a mass $>3.2~M_\odot$, above $M_{\rm max}$ for any reasonable NS models, while a ${\cal M} \lesssim 1.2M_\odot$ suggest that the mass of both compact objects need to be $<2 M_\odot$ unless one of the NSs is smaller than $1 M_\odot$, in which case the
limit to the heavier object is $3M_\odot$ \citep{Bartos2013,Hannam2013}. Together with the expected detection rate of the GRBs with GW signals (see Sec.\ref{sec:SGRB-prospect}), we think that in the era of the GW astronomy, reasonably large BNS merger GRB samples and NS-BH merger GRBs sample will be available and our goals will be (at least partly) achievable.

\section{Measuring the velocity of the gravitational wave and testing the Einstein's equivalence principle}\label{sec:speed}
\subsection{Measuring the GW velocity: the ``canonical" approach}\label{sec:speed-1}

According to general relativity, in the limit in which the wavelength of gravitational
waves is small compared to the radius of curvature of the
background space-time, the waves propagate with the velocity of the light, i.e., $c$
 \citep[see][and the references]{Will1998}. In other theories, the speed $v_{\rm g}$
could differ from $c$.
{Let us define the parameter
\begin{equation}
\varsigma\equiv (c-v_{\rm g})/c.
\end{equation}
If the gravitational wave velocity is subluminal (i.e., $\varsigma>0$), then cosmic rays lose
their energy via gravitational Cherenkov radiation significantly. The detection of ultra-high energy cosmic rays thus imposes a stringent constraint $0\leq \varsigma\leq 2\times 10^{-15}~({\rm or~ even~2\times 10^{-19}})$ (i.e., the ``subluminal constraint"), depending on the Galactic or extragalactic origin of such particles \citep{Caves1980,Moore2001}. However, there is no theoretical argument (or pathology) against GWs propagating faster than light \citep[see][and the references therein]{Nishizawa2014,Blas2016} and the weak bounds from radiation damping in binary systems are $\varsigma>-0.01$ \citep{Yagi2013}. The time-lag of arrival times between the GW and the simultaneously radiated photons is
\begin{eqnarray}
\delta t_{\rm o} &=& {1\over c}\int^{z_{\rm o}}_{0}(1+z)\left({c\over v_{\rm g}}-1\right) {\rm d}l \nonumber\\
                 &=& {1\over H_0}\int^{z_{\rm o}}_{0}\left({\varsigma \over 1-\varsigma}\right) {{\rm d}t\over
                      \sqrt{\Omega_{\rm M}(1+z)^{3}+\Omega_{\Lambda}}},
                      \label{eq:t_o}
\end{eqnarray}
where ${\rm d}l=cdz/[(1+z)H_0\sqrt{\Omega_{\rm M}(1+z)^{3}+\Omega_{\Lambda}}]$ is the differential distance the photons have traveled. Note that in this work we take the flat cosmological model (i.e., $\Omega_{\rm M}+\Omega_{\Lambda}=1$), $\Omega_{\rm M}=0.315$ and $H_0\approx 70~{\rm km~s^{-1}~Mpc^{-1}}$ is the Hubble's constant \citep{Planck2014,Riess2011}. In general, $\varsigma$ may be a function of the GW frequency ($f$) and especially when graviton mass is non-zero (i.e., $m_{\rm g}>0$) that gives $\varsigma \approx m_{\rm g}^{2}c^{4}/2h^{2}f^{2}$, where $h$ is the Planck's constant. For simplicity we assume a constant $\varsigma$ and focus on the GW/electromagnetic counterpart association at redshifts $z\ll 1$. Hence eq.(\ref{eq:t_o}) yields \citep[see also][]{Will1998,Nishizawa2014}
\begin{equation}
\varsigma \approx 5\times 10^{-17}~\left({200 ~{\rm Mpc} \over D}\right)\left({\delta t_{\rm o}\over 1~{\rm s}}\right).
\label{eq:constraint-1}
\end{equation}
In reality usually the photons and the coalescence are not simultaneous and we have $\delta t_{\rm o}=(1+z)\Delta t_{\rm e}-\Delta t_{\rm GW-ph}$, where $\Delta t_{\rm GW-ph}$ and $\Delta t_{\rm e}$ are the differences in arrival time and emission time, respectively, of the GW and the photons. In most cases, it is rather hard to get a priori value for $\Delta t_{\rm e}$. Assuming $\Delta t_{\rm e}=0$ (i.e., the GW and the electromagnetic counterparts were emitted simultaneously; see \citet{Nishizawa2016} for a more general discussion), we constrain the absolute amplitude of $\varsigma$  as
\begin{equation}
|\varsigma|<5\times 10^{-17}~\left({200 ~{\rm Mpc} \over D}\right)\left({\Delta t_{\rm GW-ph}\over 1~{\rm s}}\right).
\label{eq:constraint-2}
\end{equation}
We call the above process as the ``canonical approach" of measuring the GW velocity directly, in which the graviton and photon are assumed to have the same journey (i.e., the Einstein's equivalence principle (EEP) is guaranteed). The advantage is that as long as a GW/GRB association is established one can constrain $|\varsigma|$ directly. In Sec.\ref{sec:EEP} we outline an approach of measuring the GW velocity with a simultaneous test of EEP.
}

Some widely discussed electromagnetic counterparts of compact object mergers include \citep{Metzger2012}: (a) the (short) gamma-ray bursts and X-ray flares; (b) the afterglow emission of the (off-beam) gamma-ray burst outflows; (c) the macronova/kilonova emission of the sub-relativistic r-process material ejected during the merger; (d) the radio radiation of the forward shock driven by the sub-relativistic outflow launched during the merger. These scenarios hold for both NS-NS and NS-BH mergers (please note that for systems with very massive BHs, the NSs would be swallowed entirely and no bright electromagnetic counterparts are expected). {\it To constrain $|\varsigma|$ (see eq.(\ref{eq:constraint-2})), the time delay between the merger and the ``emergence" of the electromagnetic counterpart (i.e., $\Delta t_{\rm GW-GRB}$) is needed.}

The intense gamma-ray emission is expected to be within seconds after the merger (see Table \ref{tab:Summary} for the model-dependent estimate, where $\Delta t_{\rm GW-GRB}$ is the same as the $\Delta t_{\rm GW-ph}$ needed in eq.(\ref{eq:constraint-2})). The X-ray flares may appear within tens seconds and sometimes may last $\sim 10^{3}$ s or even longer. The challenge of detecting the ``orphan" soft X-ray signal is the lack of X-ray detector(s) with a wide field of view until the successful performance of Einstein Probe \citep[http://ep.bao.ac.cn;][]{Yuan2014} after 2022. Since advanced LIGO/VIRGO are expected in full run in 2019, here we just focus on the detectors that may (still) work at that time and hence will {\it not} discuss the X-ray signal any longer.

\begin{figure}
\begin{center}
\includegraphics[width=0.5\textwidth]{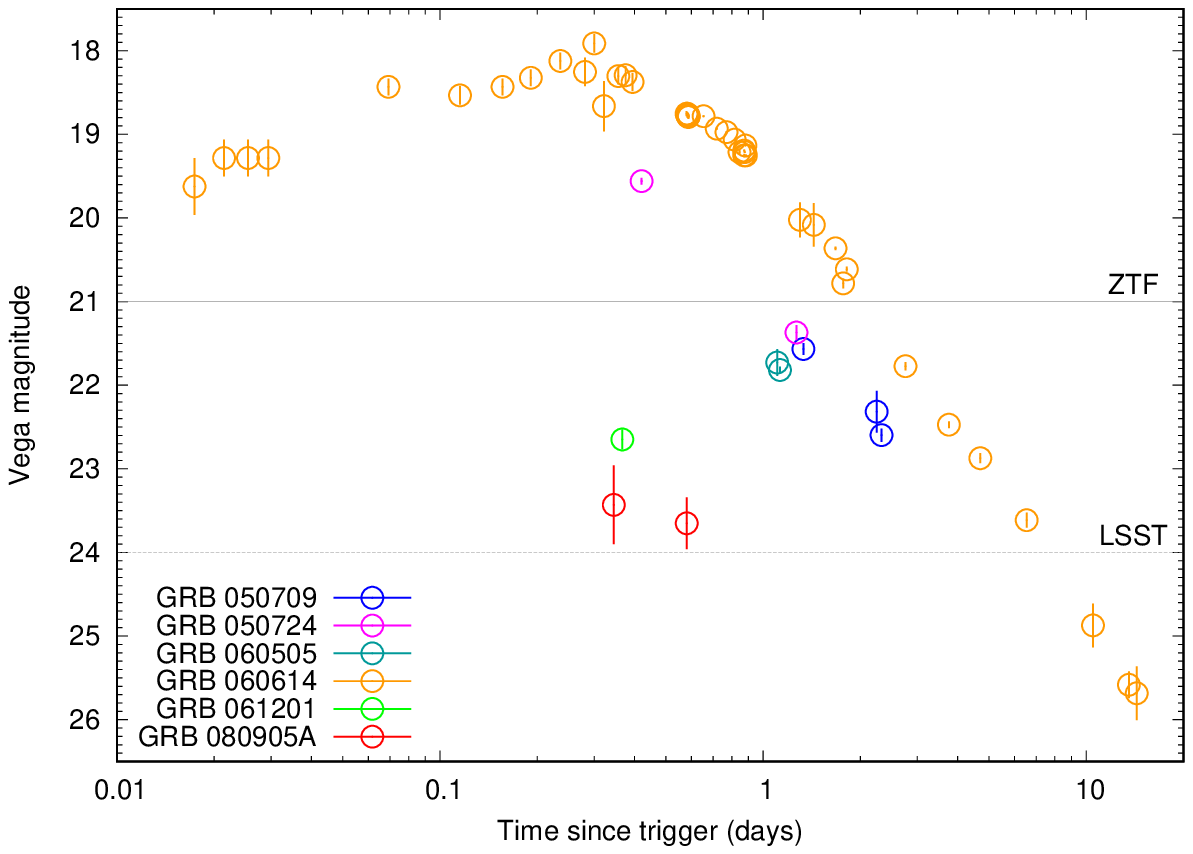}
\end{center}
\caption{The {\it r}-band afterglow emission of nearby SGRBs and long-GRBs if they took place at a luminosity distance $\sim 400$ Mpc. The initial data are adopted from \citet{Covino2006}, \citet{Malesani2007}, \citet{Ofek2007}, \citet{Xu2009}, and \citet{Fong2015}.}
\label{fig:2}
\end{figure}

If the ultra-relativistic outflow is ``on-beam", the optical/radio afterglow are relatively long-lasting and the peak of the forward shock optical
emission is expected to be within $10^{2}-10^{3}$ seconds after the merger, mainly depending on the initial bulk Lorentz factor of the outflow.  The optical afterglow emission of some on-beam GRBs, but missed by the gamma-ray detector(s), are expected to be detectable for ZTF and LSST \footnote{ In 2017 the Zwicky Transient Facility
  (ZTF) with an instantaneous field of view $\sim470$ degrees and an {\it r}-band
sensitivity $\sim $21th mag will have first light at Palomar Observatory
(http://www.ptf.caltech.edu/ztf). In one full night the
survey field of view is expected to be $\sim2.4\times10^{4}$ square degrees,
almost half of the sky. The Large Synoptic Survey Telescope (LSST) with a 9.6 deg$^{2}$ field of view that can image about
10,000 square degrees of sky in three clear nights down to limit of $\sim 24$th magnitude (Vega system) in {\it r}-band are expected to play an important role in detecting the nearby GRB afterglow and even the macronovae \citep{LSST2008}.} if the bursts are within the sensitivity distance of advanced LIGO/VIRGO network (see Fig.\ref{fig:2}).  We do not discuss the radio afterglow from SGRBs and long-short GRBs since they were rarely detected \citep[see][and the references]{Fong2015}.  If the GRB outflow is ``off-beam" with an angle separation $\Delta \theta$, the forward shock emission won't ``enter" the line of sight until its bulk Lorentz factor has dropped to $\approx 1/\Delta \theta$ \citep{Janka2006}. The off-beam timescale is related to the on-beam one as $dt_{\rm off}\approx (1+\Gamma^2\Delta\theta^{2})dt_{\rm on}$. On the other hand, we have $\Gamma \approx 7~E_{\rm k,51}^{1/8}n_{-2}^{-1/8}t_{\rm on, d}^{-3/8}$, where $E_{\rm k,51}$ is the kinetic energy of the GRB outflow in unit of $10^{51}~{\rm erg~s^{-1}}$, $n_{-2}$ is the number density of the circum burst medium in unit of $10^{-2}~{\rm cm}^{-3}$ and $t_{\rm d}$ is the timescale in unit of day. Hence the peak emission time of the ``off-beam" relativistic ejecta can be estimated as $t \sim 10~{\rm day}~E_{\rm k,51}^{1/3}n_{-2}^{-1/3}(\Delta \theta/0.2)^{8/3}$. At such a late time, the forward shock optical emission is likely (much) dimer than $22^{\rm th}$ mag for a source at a distance of $\sim 400$ Mpc (see Fig.\ref{fig:2})
and the detection prospect is not very promising.

The radio emission caused by the sub-relativistic outflow is expected to peak in years after the merger \citep{Nakar2011}, too late to be of our interest.  As for the macronova emission, in ultraviolet/optical band peak is likely in a few days while the infrared emission may peak in one or two weeks \citep[see e.g.][for the theoretical predictaions; Please see Jin et al. 2015 for the first observed multi-epoch/band macronova lightcurve]{Li1998,Barnes2013,Hotokezaka2013ApJL}. As {shown in \citet{Jin2016},} macronova emission are {likely} to have a very-promising detection prospect and can serve as ideal electromagnetic signal of the merger events. The typical discovery timescale is likely $\sim 1-10$ days.

\begin{figure}
\begin{center}
\includegraphics[width=0.5\textwidth]{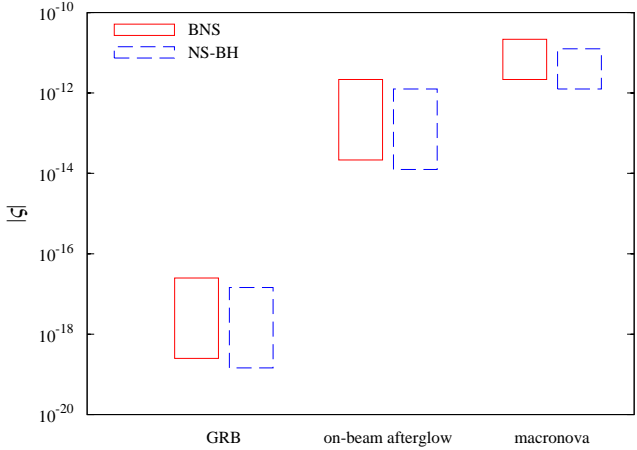}
\end{center}
\caption{Expected constraints on the difference between the GW propagation velocity and the speed of light (i.e., $|\varsigma|$) in the cases of different kinds of electromagnetic counterparts. The solid and dashed rectangles are for binary neutron star mergers and neutron star-stellar-mass black hole mergers, respectively.} \label{fig:3}
\end{figure}

We therefore conclude that if the electromagnetic counterparts are (prompt GRB emission, on-beam forward shock emission, macronova emission),
\[\Delta t_{\rm GW-ph}\sim 0.01-1~{\rm s}, ~0.01-1~{\rm day},~1-10~{\rm days},\]
respectively. The expected constraints on $\varsigma$ are shown in Fig.\ref{fig:3}. Possibly in a few years, $|\varsigma|< 10^{-18}$ is achievable.

\subsection{Measuring the GW velocity with a simultaneous test of Einstein's equivalence principle}\label{sec:EEP}
{In this subsection, we consider a simultaneous constraint on the departure of GW velocity from the speed of light and the possible violation degree of EEP with a set of GW/GRB association data.
Within the framework of parameterized post-Newtonian approximation (PPN), deviations from EEP are described by the parameter $\gamma$, which is $1$ in general relativity \citep{Will2014}. In general, the Shapiro time-delay is calculated by
$t_{\rm gra}=-{1+\gamma \over c^{3}}\int^{r_{\rm e}}_{r_{\rm o}}
U\big({r}(t); t\big)dr$,
where the integration is along the path of the photon emitted form the source at $r_{\rm e}$ and received at $r_{\rm o}$, and $U\big({r}(t); t\big)$ is the gravitational potential \citep{Shapiro1964}. If
the PPN parameter $\gamma$ are variable for different species of neutral particles, two kinds of particles emitted simultaneously from the source would arrive at different times, and the corresponding time-lag is governed by
$\Delta t_{\rm gra}=-{\Delta \gamma \over c^{3}}\int^{r_{\rm e}}_{r_{\rm o}}
U\big({r}(t); t\big)dr$ \citep{Longo1988,Krauss1988,Sivaram1999,Shapiro1964}.
In this work we focus on the Shapiro time-delay between photons and GWs caused by the gravitational
potential of the Milky Way (see Sivaram (1999) for the brief idea and Wu et al. (2016) for the dedicated investigation). Moreover, we adopt the Keplerian potential, for which the Shapiro time-delay
can be well approximated by \citep{Longo1988,Sivaram1999,Misner1973,Wu2016}
\begin{eqnarray}
\Delta t_{\rm gra} &\approx & \Delta \gamma {GM_{\rm MW}\over c^{3}}\ln \left({D\over b}\right)\nonumber\\
&\approx & 1.7\times 10^{7}~{\rm s}~ \Delta \gamma \left({M_{\rm MW}\over 6\times 10^{11}M_\odot}\right){\ln ({D/b})\over 4\ln10},
\label{eq:Shapiro}
\end{eqnarray}
where $M_{\rm MW}\approx 6\times 10^{11}~M_\odot$ is the mass of Milky Way, $D$ is the distance of the cosmological transient to the Earth, and $b$ is the impact parameter of the particle paths relative to the Milky Way center, {and we have normalized $\ln (D/b)$ to the value of $4\ln 10$ to address the facts that $D \sim 100$ Mpc in the advanced LIGO/VIRGO era and $d \sim 10$ kpc}.

As in the ``canonical approach", we assume that $v_{\rm g}$ is a constant. Then the observed time delay consists of three parts, i.e.,
\begin{equation}
\Delta t_{\rm GW-ph}=\Delta t_{\rm e}-\delta t_{\rm o}+\Delta t_{\rm gra}.
\label{eq:self-cons}
\end{equation}
where $\delta t_{\rm o} \approx 2\times 10^{16}~{\rm s}~\varsigma (D/200~{\rm Mpc})$. In the specific model of ``Dark Matter Emulators" the GWs are expected to arrive earlier than the simultaneously-emitted GRB photons by $\sim 10^{3}$ days \citep{Desai2008}. While in reasonable astrophysical models the GW signal should precede the GRB for a given source. Hence an almost simultaneous arrival of the GW/GRB signals requires a subluminal GW with a $\varsigma \sim 4\times 10^{-9}(D/200~{\rm Mpc})$, which violates the ``submuminal constraint" set by the ultra-high energy cosmic rays (i.e., $0<\varsigma<2\times 10^{-15}$) and in turn rules out the Dark Matter Emulators but favors the dark matter model (see also \citet{Kahya2016} for the discussion on the possible GW150914/GBM transient association).

If two events of GW/electromagnetic counterpart association are observed (which are marked by subscript 1 and 2, respectively), we have
\begin{equation}
\Delta \gamma =5.5\times 10^{-7}\left({M_{\rm MW}\over 6\times 10^{11}~M_\odot}\right)^{-1}{ D_{2} \Delta t'_{\rm GW-ph,1}-D_{1}\Delta t'_{\rm GW-ph,2}\over D_{2}{\ln ({D_{1}\over b_{1}})}-D_{1}{\ln ({D_{2}\over b_{2}})}},
\end{equation}
and
\begin{equation}
\varsigma=10^{-16}{100~{\rm Mpc}~\left[\ln({D_{2}\over b_{2}}) \Delta t'_{\rm GW-ph,1}-\ln({D_{1}\over b_{1}})\Delta t'_{\rm GW-ph,2}\right]\over D_{2}\ln (D_{1}/b_{1})-D_{1}\ln (D_{2}/b_{2})},
\end{equation}
where $\Delta t'_{\rm GW-ph,1}\equiv \Delta t_{\rm GW-ph,1}-\Delta t_{\rm e,1}$ and $\Delta t'_{\rm GW-ph,2}\equiv \Delta t_{\rm GW-ph,2}-\Delta t_{\rm e,2}$ (Note that the relationship $\Delta t_{\rm GW-ph}\geq \Delta t_{\rm e}$, as expected in general relativity, may be invalid for current model assumption).  For a set of GW signals with electromagnetic counterparts, the $(\Delta t_{\rm GW-ph},~D,~b)$ are available and the main uncertainty on constraining $\varsigma$ and $\Delta \gamma$ are the accuracy of estimating $\Delta t_{\rm e}$.
Here we do not simply take $\Delta t_{\rm e}$ as zero and focus on the events of SGRBs or short-long GRBs with associated GW signals, {for which $\Delta t_{\rm e}$ can be relatively reasonably estimated (see Table 1). In particular, for NS-BH merger GRBs we expect that $\Delta t_{\rm e}\leq T_{90}$ (unless $T_{90}$ is significantly shorter than $\Delta t_{\rm laun}~(\sim 10~{\rm ms})$ that has not been recorded by {\it Swift} yet). }

The conservative constraints on $|\Delta \gamma |$ and $|\varsigma|$ are
\begin{eqnarray}
|\Delta \gamma | &\leq& 5.5\times 10^{-7}\left({M_{\rm MW}\over 6\times 10^{11}~M_\odot}\right)^{-1}\nonumber\\
&&{| D_{2} \Delta t_{\rm GW-GRB,1}-D_{1}\Delta t_{\rm GW-GRB,2}|+
| D_{2} \Delta t_{\rm e,1}-D_{1}\Delta t_{\rm e,2}|
\over |D_{2}{\ln (D_{1}/b_{1})}-D_{1}{\ln (D_{2}/b_{2})}|},\nonumber\\
\label{eq:self-cons-1}
\end{eqnarray}
and
\begin{eqnarray}
|\varsigma| &\leq & 10^{-16}\nonumber\\
&&\ln({D_{2}\over b_{2}}){| \Delta t_{\rm GW-GRB,1}-{\cal R}\Delta t_{\rm GW-GRB,2}|
+|\Delta t_{\rm e,1}-{\cal R}\Delta t_{\rm e,2}|
\over |{D_{2}\over 100~{\rm Mpc}}\ln (D_{1}/b_{1})-{D_{1}\over 100~{\rm Mpc}}\ln (D_{2}/b_{2})|},\nonumber\\
\label{eq:self-cons-2}
\end{eqnarray}
respectively, where ${\cal R}\equiv \ln(D_{1}/b_{1})/\ln({D_{2}/b_{2}})$. {If in the future data people can identify two or more NS-BH merger GRBs associated with GW signals, as a {\it conservative} estimate we take into account the fact that $|\Delta t_{\rm e,1}-{\cal R}\Delta t_{\rm e,2}|< \Delta t_{\rm e,1}+{\cal R}\Delta t_{\rm e,2}$ and further replace $\Delta t_{\rm e}$ by $T_{90}$. For $D_{2}\geq 1.5D_{1}$, $b_{2}\sim b_{1}$, $\Delta t_{\rm GW-GRB,1}\approx \Delta t_{\rm GW-GRB,2}$ and $T_{90,1} \approx T_{90,2}$,} the above two constraints (on $|\Delta \gamma |$ and $|\varsigma|$) further reduce to
\begin{equation}
|\Delta \gamma |< 6\times 10^{-8}~\left({M_{\rm MW}\over 6\times 10^{11}~M_\odot}\right)^{-1}{\Delta t_{\rm GW-GRB,1}+T_{\rm 90,1}\over [\ln(D_1/b_1)/4\ln10]},
\end{equation}
and
\begin{equation}
|\varsigma|< 10^{-17}~{(\Delta t_{\rm GW-GRB,1}+T_{\rm 90,1})\over [\ln(D_1/b_1)/4\ln10]}{\ln(D_{2}/D_{1})\over [(D_2-D_1)/100~{\rm Mpc}]}.
\end{equation}
Interestingly, such constraints can be as tight as the bounds on $\Delta \gamma$ or $|\varsigma|$ set by excluding either the EEP test or the departure of $v_{\rm g}$ from $c$.}

\section{Discussion}
SGRBs are widely believed to be powered by the mergers of compact object binaries. Note that the BH$-$NS merger rate is generally expected to be $\sim 1/10$ times that of the NS$-$NS merger rate \citep{Abadie2010}. Hence most SGRBs are expected to be from NS$-$NS mergers and a small fraction of events may be due to NS$-$BH mergers. Though in the upcoming era of advanced LIGO/VIRGO network the prospect of detecting GW-associated SGRBs is promising, none of the nearby (i.e., $z<0.2$) SGRBs are found within the sensitivity distance of the upcoming advanced LIGO/VIRGO network $D_{\rm *,BNS}\approx 400~{\rm Mpc}~(9/\rho_*)$ for $\rho_*\geq 9$. Such a non-identification, though still understandable (see eq.(\ref{eq:D-2})), is somewhat disappointing. Interestingly, we find out that GRB 060505, one supernova-less long event (also known as long-short GRB), if powered by a NS-NS merger, is likely within the distance of $D_{\rm *,BNS}(\rho_*\approx 9)$. The other long-short burst GRB 060614, accompanied with a macronova signal that is plausibly powered by a NS$-$BH merger, is within the distance of $D_{\rm *,NS-BH}(\rho_*\approx 9)$.
Therefore in the era of GW astronomy, the compact object binary merger origin of some long-short GRBs, as favored by the macronova signature displayed in GRB 060614, will be unambiguously tested. We hence suggest that both SGRBs and long-short GRBs are prime targets of the advanced LIGO/VIRGO network and $\gamma-$ray detectors with wide field of views are encouraged to monitor the sky continually to get the accurate information of the prompt emission properties.

In the era of the advanced LIGO/VIRGO, reasonably large BNS merger GRB samples and NS-BH merger GRB samples are establishable (see Sec.2). Motivated by such a fact, we have examined the possible distribution of $\Delta t_{\rm GW-GRB}$ and the relation between $T_{90}$ and $\Delta t_{\rm GW-GRB}$ for each sample. As summarized in Tab. 1, in the case of $R_{\rm pro}\ll 10^{16}$ cm that represents the scenarios of photospheric radiation and regular (magnetized) internal shock radiation, it is expected that (1) for BNS mergers $\Delta t_{\rm GW-GRB}$ is significantly longer than that of the NS-BH mergers, i.e., $\Delta t_{\rm GW-GRB}({\rm NS-BH})\ll \Delta t_{\rm GW-GRB}({\rm NS-NS})$; (2) for NS-BH mergers, usually $\Delta t_{\rm GW-GRB}$ is expected to be shorter than $T_{90}$, i.e., $\Delta t_{\rm GW-GRB}({\rm NS-BH})<T_{90}$. While for $R_{\rm pro}\sim 10^{16}~{\rm cm}$, we expect that $\Delta t_{\rm GW-GRB}$ should be comparable with $T_{90}$ for both BNS and NS-BH merger powered SGRBs. The comparison with future real data will be helpful in revealing the central engine physics. {We would like to also point out that in some specific astrophysical or new physics scenarios,  the GW may precede the GRB signal significantly (i.e., $\Delta t_{\rm GW-GRB}>10$ s). If such large time-lags indeed presents in the future data, the statistical study of the distribution of $\Delta t_{\rm GW-GRB}>10~{\rm s}$ can distinguish between the astrophysical model (for example the model proposed in \citet{Rezzolla2015} for some BNS mergers but not for NS-BH mergers) and the new physics (e.g., the supluminal movement of the GW).}

To tightly constrain the difference between the GW velocity and the speed of light, the shorter $\Delta t_{\rm GW-ph}$ the better (see Sec.\ref{sec:speed}). If the electromagnetic counterpart of GW signal is GRB, we have $\Delta t_{\rm GW-ph}=\Delta t_{\rm GW-GRB}$. The shortest $\Delta t_{\rm GW-GRB}$ is expected for the prompt BH formation in the NS-NS mergers or the NS-BH mergers if the onset of the prompt GRB emission is governed by the photosphere or regular internal shocks (i.e., $R_{\rm pro}\ll 10^{16}$ cm), in such a case the constraint $|\varsigma|<10^{-16}$ or even tighter is possible (If the GW150914/GBM transient 150914 association is intrinsic, $|\varsigma|<10^{-17}$ is inferred, as shown in \citet{Li2016} and \citet{Ellis2016}). {With two GW/GRB association events that are expected to be available in the near future we can measure the GW velocity with a simultaneous test of EEP. Intriguingly, in such treatments very accurate measurements/tests are still achievable (see Sec.\ref{sec:EEP}).}

\section*{Acknowledgments}
We thank the anonymous referee for helpful suggestions, Tsvi Piran for the discussion, Chris Messenger and Imre Bartos for the comments. This work was supported in part by National Basic Research Programme of China (No. 2013CB837000 and No. 2014CB845800),
NSFC under grants 11525313 (i.e., Funds for Distinguished Young Scholars), 11361140349, 11273063 and 11433009,
the Foundation for Distinguished Young Scholars of Jiangsu Province, China (Grant No. BK2012047)
and the Strategic Priority Research Program (Grant No. XDB09000000).\\

\clearpage


\begin{thebibliography}{}

\bibitem[Aasi et al. (2013)]{Aasi2013} Aasi, J., Abadie, J., Abbott, B. P., et al. (LIGO Scientific Collaboration, Virgo Collaboration) 2013, arXiv:1304.0670

\bibitem[Aasi et al. (2014a)]{Aasi2014a} Aasi, J., Abadie, J., Abbott, B. P., et al. (LIGO Scientific Collaboration, Virgo Collaboration) 2014, Phys. Rev. D 89, 122004

\bibitem[Aasi et al. (2014b)]{Aasi2014b} Aasi, J., Abadie, J., Abbott, B. P., et al. (LIGO Scientific Collaboration, Virgo Collaboration) 2014, Phys. Rev. Lett. 113, 011102

\bibitem[Abbott et al. (2016)]{Abbott2016} Abbott, B. P. et al. (LIGO Scientific Collaboration, Virgo Collaboration), 2016, Phys. Rev. Lett., 116, 061102

\bibitem[Abadie et al. (2010)]{Abadie2010} Abadie, J., Abadie, J., Abbott, B. P., et al. 2010, CQGra, 27, 173001

\bibitem[Abadie et al. (2012)]{Abadie2012} J. Abadie, B. P. Abbott, R. Abbott, et al. (LIGO Scientific Collaboration, Virgo Collaboration) 2012, ApJ., Volume 760, Number 1

\bibitem[Abadie et al. (2015)]{Abadie2015} Abadie, J., Abadie, J., Abbott, B. P., et al. 2015, CQGra, 32, 074001

\bibitem[Acernese et al. (2015)]{Acernese2015} F. Acernese, et al., 2015, CQGra, 32, 024001

\bibitem[Ackermann et al. (2010)]{Ackermann2010} Ackermann, M., et al., 2010, ApJ, 716, 1178

\bibitem[Amati et al. (2007)]{Amati2007} Amati, L., et al. 2007, A\&A, 463, 913

\bibitem[Antoniadis et al. (2013)]{Antoniadis2013} Antoniadis, J., et al. 2013, Science, 340, 1233232

\bibitem[Bailyn et al. (1998)]{Bailyn1998} Bailyn, C. D., Jain, R. K., Coppi, P.,
\& Orosz, J. A. 1998, ApJ, 499, 367

\bibitem[Balbus \& Hawley (1991)] {Balbus1991} Balbus, S. A., \& Hawley, J. F., 1991, ApJ, {376}, 214
(1991).

\bibitem[Barnes \& Kasen (2013)]{Barnes2013} Barnes, J. \& Kasen, D. 2013, ApJ, 773, 18.

\bibitem[Bartos et al. (2013)]{Bartos2013} Bartos, I., Brady, P., \& M\'{a}rka, M., 2013, Class. Quantum Grav. 30, 123001

\bibitem[Bartos \& Marka (2015)]{Bartos2015} Bartos, I., \&  Marka, S. 2015, Phys. Rev. Lett. 115, 231101

\bibitem[Baumgarte et al. (2000)]{Baumgarte2000} Baumgarte, T. W., Shapiro, S. L., \& Shibata, M. 2000, ApJL, 528, L29

\bibitem[Belczynski et al. (2010)]{Belczynski2010}
 Belczynski, K., Dominik, M., Bulik, T.,  O'Shaughnessy, R., Fryer, C., \& Holz, D. E.
 2010, ApJL, 715, L138

\bibitem[Belczynski et al. (2016)]{Belczynski2016}
 Belczynski, K., Holz, D. E., Bulik, T., O'Shaughnessy, R. 2016, arXiv:1602.04531

\bibitem[Berger et al. (2013)]{Berger2013} Berger, E., Fong, W., \& Chornock, R. 2013, ApJL, 744, L23

\bibitem[Berger (2014)]{Berger2014} Berger, E., 2014, ARA\&A, 52, 43

\bibitem[Blandford \& Znajek (1977)]{Blandford1977} Blandford, R. D., \& Znajek, R. L. 1977, MNRAS, 179, 433

\bibitem[Blas et al. (2016)]{Blas2016} Blas, D, Ivanov, M. M., Sawicki, I., \& Sibiryakov, S. 2016, arXiv:1602.04188

\bibitem[Bromberg \& Tchekhovskoy (2015)]{Bromberg2015} Bromberg, O., \& Tchekhovskoy, A., 2015, arXiv:1508.02721

\bibitem[Bromberg et al. (2014)]{Bromberg2014} Bromberg, O., Granot, J., Lyubarsky, Y., \& Piran, T. 2014, MNRAS, 443, 1532



\bibitem[Caves (1980)]{Caves1980} Caves, C. M. 1980, Ann. Phys., 125, 35

\bibitem[Charisi et al. (2015)]{Charisi2015} Charisi, M. M\'{a}rka, S., \& Bartos, I. 2015, MNRAS,  448, 2624

\bibitem[Chhotray \& Lazzati (2015)]{Chhotray2015} Chhotray, A., \& Lazzati, D., 2015, ApJ, 802, 132

\bibitem[Clark et al. (2015)]{Clark2015} Clark, J. et al. 2015, ApJ, 809, 53

\bibitem[Connaughton et al. (2016)]{Connaughton2016} Connaughton, V., et al. 2016, arXiv:1602.03920

\bibitem[Covino et al. (2006)]{Covino2006} Covino, S., et al. 2006, A\&A, 447, L5

\bibitem[Dai et al. (2006)]{Dai2006} Dai, Z. G., Wang, X. Y., Wu, X. F., \& Zhang, B., 2006, Science, 311, 1127

\bibitem[Daigne \& Mochkovitch (2002)]{Daigne2002} Daigne, F., \& Mochkovitch, R. 2002, MNRAS, 336, 1271

\bibitem[D'Avanzo et al. (2014)]{DAvanzo2014} D'Avanzo, P. et al. 2014, MNRAS, 442, 2342

\bibitem[Della Valle et al. (2006)]{DellaValle2006} Della Valle, M., Chincarini, G., Panagia, N.,  et al. 2006, Natur, 444, 1050

\bibitem[Desai et al. (2008)]{Desai2008} Desai, S., Kahya,  E. O., \&  Woodard, R. P. 2008, Phys. Rev.
D 77, 124041

\bibitem[Dietz et al. (2013)]{Dietz2013}  Dietz, A.., Fotopoulos, N.,  Singer, L., \& Cutler, C. 2013, Phys.
Rev. D, 87, 064033

\bibitem[Eichler et al. (1989)]{Eichler1989} Eichler D.,  Livio M.,  Piran T., \& Schramm D. N. 1989, Natur, 340, 126

\bibitem[Ellis et al. (2016)]{Ellis2016} Ellis, J. et al. 2016, arXiv:1602.04764


\bibitem[Faber \& Rasio (2012)]{Faber2013} Faber, J. A., \& Rasio, F. A.  2012, Living Rev. Relativity, 15, 8

\bibitem[Fairhurst (2011)]{Fairhurst2011} Fairhurst, S., 2011,
CQGra, 28, 105021

\bibitem[Fan \& Wei (2011)]{Fan2011} Fan, Y. Z., \& Wei, D. M. 2011, ApJ, 739, 47

\bibitem[Fan et al. (2004)]{Fan2004} Fan, Y. Z., Wei, D. M., \& Zhang, B. 2004, MNRAS, 354, 1031

\bibitem[Fan et al. (2012)]{Fan2012} Fan, Y. Z., Wei, D. M., Zhang, F. W., \& Zhang, B. B. 2012, ApJL, 755, L6

\bibitem[Fan et al. (2005)]{Fan2005} Fan, Y. Z., Zhang, B., \& Proga, D. 2005, ApJL, 635, L129

\bibitem[Farr et al. (2011)]{Farr2011} Farr, W. M., et al. 2011, ApJ, 741, 103

\bibitem[Finn et al. (1999)]{Finn1999} Finn, L. S.,  Mohanty, S. D., \& Romano, J. D., 1999, Phys. Rev. D
60, 121101

\bibitem[Finn \& Sutton (2002)]{Finn2002} Finn, L. S., \& Sutton, P. J., 2002,  Phys.
Rev. D, 65, 044022



\bibitem[Fong et al. (2015)]{Fong2015} Fong, W. et al., 2015, ApJ, 815, 102

\bibitem[Fox et al. (2005)]{Fox2005} Fox, D. B., et al. 2005, Nature, 437, 845

\bibitem[Foucart et al. (2014)]{Foucart2014} Foucart, F., et al. 2014, Phys. Rev. D., arXiv:1405.1121

\bibitem[Foucart et al. (2015)]{Foucart2015} Foucart, F., et al. 2015, Phys. Rev. D., arXiv:1502.04146

\bibitem[Fynbo et al. (2006)]{Fynbo2006} Fynbo, J. P. U., Watson, D., Th\"one, C. C., et al. 2006, Natur, 444, 1047

\bibitem[Gal-Yam et al. (2006)]{Gal-Yam2006} Gal-Yam, A., Fox, D. B., Price, P. A., et al. 2006, Natur, 444, 1053

\bibitem[Galama et al. (1998)]{Galama1998} Galama, T. J. et al. 1998, Nature 395, 670

\bibitem[Gehrels et al. (2004)]{Gehrels2004} Gehrels, N., et al. 2004, ApJ, 611, 1005

\bibitem[Gehrels et al. (2006)]{Gehrels2006} Gehrels, N., Norris, J. P., Barthelmy, S. D., et al. 2006, Natur, 444, 1044


\bibitem[Giannios (2008)]{Giannios2008} Giannios, D., 2008,  A\&A, 480, 305


\bibitem[Granot et al. (2015)]{Granot2015} Granot, J., Piran, T., Bromberg, O., Racusin, J., Judith, L., \& Daigne, F., 2005, arXiv:1507.08671


\bibitem[Hannam et al. (2013)]{Hannam2013}
 Hannam, M., Brown, D. A., Fairhurst, S., Fryer, C. L., \& Harry, I. W. 2013, ApJL, 766, L14

\bibitem[Harry \& Fairhurst (2011)]{Harry2011} Harry, I. W.  \&  Fairhurst, S. 2011, Phys. Rev. D, 83, 084002

\bibitem[Hotokezaka et al. (2013a)]{Hotokezaka2013} Hotokezaka, K. et al., 2013, {Phys. Rev. D.},  {87}, 024001

\bibitem[Hotokezaka et al. (2013b)]{Hotokezaka2013ApJL} Hotokezaka, K.,  Kyutoku, K., Tanaka, M., et al. 2013, ApJL, 778, L16

\bibitem[Ivezi\'{c} et al. (2008)]{LSST2008} Ivezi\'{c}, Z., et al. LSST: from Science Drivers to Reference Design and Anticipated Data Products. arXiv:0805.2366

\bibitem[Janka et al. (1999)]{Janka1998} Janka, H. T.,  Eberl, T., R. Maximilian, \& Fryer, C. L. 1999, ApJ, 527, L39

\bibitem[Janka et al. (2006)]{Janka2006} Janka, H. T., Aloy, M. A., Mazzali, P. A., \& Pian, E. 2006, ApJ, 645, 1305

\bibitem[Jin et al. (2015)]{Jin2015} Jin, Z. P., et al. 2015, ApJL, 811, L22

\bibitem[Jin et al. (2016)]{Jin2016} Jin, Z. P., et al. 2016, Nat. Commun. submitted (arXiv:1603.07869)

\bibitem[Kahya \& Desai (2016)]{Kahya2016} Kahya, E. O., \& Desai, S. 2016, arXiv:1602.04779

\bibitem[Kasen et al. (2013)]{Kasen2013} Kasen, D., Badnell, N. R. \& Barnes, J. 2013, ApJ, 774, 25

\bibitem[Kelley et al. (2013)]{Kelley2013} Kelley, L. Z., Mandel, I., \& Ramirez-Ruiz, E., 2013, Phys. Rev. D., 87, 123004

\bibitem[Kiuchi et al. (2015)]{Kiuchi2015} Kiuchi, K., et al. 2015, Phys. Rev. D., arXiv:1506.06811


\bibitem[Kluz\'aniak \& Ruderman (1998)]{Kluzaniak1998} Kluz\'aniak, W.  \& Ruderman, M., 1998, ApJL, 505, L113


\bibitem[Kochanek \& Piran (1993)]{Kochanek1993} Kochanek, C. S., \& Piran, T. 1993, ApJL, 417, L17

\bibitem[Kouveliotou et al. (1993)]{Kouveliotou1993} Kouveliotou, C., C. A. Meegan, G. J. Fishman, N. P. Bhat, M.
S. Briggs, T. M. Koshut, W. S. Paciesas, and G. N. Pendleton,
1993, ApJL, 413, L101

\bibitem[{Krauss \& Tremaine} (1988)]{Krauss1988} Krauss, L. M., and Tremaine, S.
1988, Phys. Rev. Lett., 60, 176

\bibitem[Kulkarni (2005)]{Kulkarni2005} Kulkarni, S. R. 2005, arXiv:astro-ph/0510256

\bibitem[Kumar \& Zhang (2015)]{Kumar2015} Kumar, P., \& Zhang, B., 2015, PhR, 561, 1



\bibitem[Lattimer (2012)]{Lattimer2012} Lattimer, J. M. 2012, Annu. Rev. Nucl. Part. Sci., 62, 485

\bibitem[Lee \& Ramirez-Ruiz (2007)]{Lee2007} Lee, W. H.  \&  Ramirez-Ruiz, E. 2007, New J. Phys. 9, 17

\bibitem[Levan et al. (2015)]{Levan2015} Levan, A. J., Hjorth, J., Wiersema, K., Tanvir, N. R. 2015, GCN Circ. 17281
(http://gcn.gsfc.nasa.gov/gcn3/17281.gcn3)

\bibitem[Li \& Paczynski (1998)]{Li1998} Li, L.-X., \& Paczy\'{n}ski, B. 1998, ApJL, 507, L59

\bibitem[Li et al. (2016)]{Li2016} Li, X., Zhang, F. W., \& Yuan, Q., et al. 2016, ApJL, 827, L16 (arXiv:1602.04460)

\bibitem[Liang et al. (2015)]{Liang2015} Liang, E.-W. et al. 2015, ApJ, 813, 116


\bibitem[Liu et al. (2015)]{Liu2015} Liu, T., Lin, Y. Q., Hou, S. J., \&  Gu, W. M. 2015, ApJ, 806, 58

\bibitem[{Longo} (1988)]{Longo1988} Longo, M. J.
1988, Phys. Rev. Lett., 60, 173

\bibitem[L\"{u} et al. (2012)]{Lv2012} L\"{u}, J., Zou, Y.-C., Lei, W.-H., et al. 2012, ApJ, 751, 49

\bibitem[Lyutikov \& Blandford (2003)]{Lyutikov2003} Lyutikov, M., \& Blandford, R. D. 2003, astro-ph/0312347


\bibitem[Meegan et al. (2009)]{Meegan2009} Meegan, C., Lichti, G., \& Bhat, P., et al. 2009, ApJ, 702, 791

\bibitem[Malesani et al. (2007)]{Malesani2007} Malesani, D., et al. 2007, A\&A, 473, 77

\bibitem[M\'{e}sz\'{a}ros et al. (1993)]{Meszaros1993} M\'{e}sz\'{a}ros, P., Laguna, P., \& Rees, M. J. 1993, ApJ, 415, 181

\bibitem[Metzger \& Berger (2012)]{Metzger2012} Metzger, B. D., \& Berger, E., 2012, ApJ, 746, 48

\bibitem[Metzger et al. (2010)]{Metzger2010} Metzger, B. D., Mart\'{i}nez-Pinedo, G., Darbha, S. et al. 2010, MNRAS, 406, 2650


\bibitem[{Misner} et al. (1973)]{Misner1973} Misner, C. W.,  Thorne, K. S., \&  Wheeler, J. A. Gravitation
(Freeman, San Francisco, 1973).

\bibitem[Moore \& Nelson (2001)]{Moore2001} Moore, G. D.  and  Nelson, A. E., 2001, JHEP, 0109, 023

\bibitem[Murguia-Berthier et al. (2014)]{Murguia-Berthier2014} Murguia-Berthier, A., Montes, G., Ramirez-Ruiz, E.,  De Colle, F., \& Lee, W. H., 2014, ApJ, 788, L8

\bibitem[Nagakura et al. (2014)]{Nagakura2014} Nagakura, H., et al. 2014, ApJL, 784, L28

\bibitem[Nakar (2007)]{Nakar2007} Nakar, E. 2007, Phys. Rep., 442, 166

\bibitem[Nakar \& Piran (2011)]{Nakar2011} Nakar, E., \& Piran, T. 2011, Nature, 478, 82

\bibitem[Narayan et al. (1992)]{Narayan1992} Narayan, R., Paczynski, B., \& Piran, T. 1992,  ApJL, 395, L83

\bibitem[Narayan et al. (2001)]{Narayan2001} Narayan, R., Kumar, P., \& Piran, T. 2001,  ApJ, 557, 949

\bibitem[Nishizawa \& Nakamura (2014)]{Nishizawa2014} Nishizawa, A., \& Nakamura, T., Phys. Rev. D, 90, 044048

\bibitem[Nishizawa (2016)]{Nishizawa2016} Nishizawa, A. 206, arXiv:1601.01072

\bibitem[Nissanke et al. (2013)]{Nissanke2013} Nissanke, S., Kasliwal, M., \& Georgieva, A., 2013, ApJ, 767, 124


\bibitem[Ozel et al. (2010)]{Ozel2010} Ozel, F., Psaltis, D., Narayan, R., \& McClintock, J. E. 2010, ApJ, 725,
1918

\bibitem[Ofek et al. (2007)]{Ofek2007} Ofek, E. O. et al. 2007, ApJ, 662, 1129

\bibitem[Paczynski (1990)]{Paczynski1990} Paczy\'{n}ski, B. 1990, ApJ, 363, 218

\bibitem[Paul et al. (2008)]{Paul2008} Paul, J., Wei, J. Y., Basa, S., \& Zhang, S. N. 2008,
 Comptes Rendus Physique, 12, 298

\bibitem[Paschalidis et al. (2015)]{Paschalidis2015} Paschalidis, V., Ruiz, M., \& Shapiro, S. L. 2015, arXiv:1410.7392

\bibitem[Piran (1999)]{Piran1999} Piran, T., 1999, Rhys. Rep., 314, 575

\bibitem[Piran et al. (1993)]{Piran1993} Piran, T., Shemi, A., \& Narayan, R. 1993, MNRAS, 263, 861

\bibitem[P\"{u}rrer (2014)]{Purrer2014} P\"{u}rrer, M., 2014, Class. Quant. Grav., 31, 195010

\bibitem[Ade et al. (2014)]{Planck2014} Ade, P. A. R., et al. [Planck collaboration], 2014, A\&A, 571, 16

\bibitem[Popham et al. (1999)]{Popham1999} Popham, R., Woosley, S. E., \& Fryer, C. 1999, ApJ, 518, 356

\bibitem[Rees \& M\'{e}sz\'{a}ros (1994)]{Rees1994} Rees, M. J., \& M\'{e}sz\'{a}ros, P., 1994, ApJ, 430, L93

\bibitem[Rezzolla \& Kumar (2015)]{Rezzolla2015} Rezzolla, L., \& Kumar, P. 2015, ApJ,  802, 95

\bibitem[Riess et al. (2011)]{Riess2011} Riess, A. G. et al. 2011, ApJ, 730, 119

\bibitem[Rowlinson et al. (2013)]{Rowlinson2013} Rowlinson, A., O'Brien, P. T., Metzger, B. D., Tanvir, N. R., Levan, A. J.
2013, MNRAS, 430, 1061


\bibitem[Savchenko et al. (2016)]{Savchenko2016} Savchenko, V., et al. 2016, ApJL, submitted (arXiv:1602.04180)

\bibitem[Shapiro (1964)]{Shapiro1964}  Shapiro, I. I.
 1964, Phys. Rev. Lett., 13, 789.

\bibitem[{Sivaram} (1999)]{Sivaram1999} Sivaram, C.
1999, Bull. Astron. Soc. India, 27, 627




\bibitem[Siegel et al. (2013)]{Siegel2013} Siegel, D., Ciolfi, R., Harte, A. I., Rezzolla, L. 2013, Phys. Rev. D, 87, 121302

\bibitem[Sekiguchi et al. (2011)]{Sekiguchi2011} Sekiguchi, Y., Kiuchi, K., Kyutoku, K., and Shibata, M., 2011,  Phys. Rev. Lett., {107}, 051102 (2011).

\bibitem[Setiawan et al. (2006)]{Setiawan2006} Setiawan, S., Ruffert, M., \&  Janka, H.-T. 2006, A\&A, 458, 553

\bibitem[Sharpiro (2000)]{Shapiro2000} Sharpiro, S. L.  2000, Astrophys. J., {544}, 397

\bibitem[Shemi \& Piran (1990)]{Shemi1990} Shemi, A., \& Piran, T. 1990, ApJ, 365, L55

\bibitem[Shibata \& Taniguchi (2006)]{Shibata2006} Shibata, M., \& Taniguchi, K.
2006, Phys. Rev. D, 73, 064027

\bibitem[Soderberg et al. (2005)]{Soderberg2005} Soderberg, A., et al. 2005, ApJ, 627, 877



\bibitem[Tanvir et al. (2013)]{Tanvir2013} Tanvir, N. R., Levan, A. J., Fruchter, A. S. et al. 2013, Natur, 500, 547

\bibitem[Th\"{o}ne et al. (2008)]{Thone2008} Th\"{o}ne, C. C., et al. 2008, ApJ, 676, 1151

\bibitem[Usov (1994)]{Usov1994} Usov, V. V. 1994, MNRAS, 267, 1035

\bibitem[Veitch et al. (2015)]{Veitch2014} Veitch, J., et al.  Phys. Rev. D, 91, 042003 (2015).


\bibitem[Wanderman \& Piran (2015)]{Wanderman2014} Wanderman, D., \& Piran, T. 2015, MNRAS, 448, 3026

\bibitem[Will (1998)]{Will1998} Will, C. M., 1998, Phys. Rev. D., 57, 2061

\bibitem[Will (2014)]{Will2014} Will, C. M. 2014, Living Rev. Relativity, 17, 4

\bibitem[Williamson et al. (2014)]{Williamson2014} Williamson, A. R. et al. 2014, Phys. Rev. D, 90, 122004

\bibitem[Wu et al. (2016)]{Wu2016} Wu, X. F., et al. arXiv:1602.01566 (2016).

\bibitem[Xu et al. (2009)]{Xu2009} Xu, D., Starling, R. L. C., Fynbo, J. P. U. et al. 2009, ApJ, 696, 971

\bibitem[Yagi et al. (2013)]{Yagi2013} Yagi, K., Blas, D., Yunes, N., \& Barausse, E. 2014, Phys. Rev.
Lett. 112, 161101

\bibitem[Yang et al. (2015)]{Yang2015} Yang, B., Jin, Z. P., Li, X. et al. 2015, Nat. Commun., 6, 7323

\bibitem[Yuan et al. (2014)]{Yuan2014} Yuan, W. M., et al. 2014, Proceedings of Swift: 10 Years of Discovery (SWIFT 10), held 2-5 December 2014 at La Sapienza University, Rome, Italy. Online at http://pos.sissa.it/cgi-bin/reader/conf.cgi?confid=233, id.6

\bibitem[Zalamea \& Beloborodov (2011)]{Zalamea2011} Zalamea, I., \& Beloborodov, A. M. 2011, MNRAS, 410, 2302

\bibitem[Zhang (2014)]{Zhang2014} Zhang, B., 2014, Int. J. Mod. Phys. D, 23, 1430002

\bibitem[Zhang \& M\'{e}sz\'{a}ros (2002)]{Zhang2002} Zhang, B., \& M\'{e}sz\'{a}ros, P. 2002, ApJ, 581, 1236

\bibitem[Zhang \& Yan (2011)]{Zhang2011} Zhang, B., \& Yan, H. R., 2011, ApJ, 726, 90

\bibitem[Zhang et al.(2007)]{Zhang2007} Zhang, B., Zhang, B. B., Liang, E. W. et al. 2007, ApJL, 655, L25

\end{thebibliography}
\end{document}